\documentclass[12pt]{revtex4} 
\usepackage{bm}
\usepackage[dvips]{graphicx}
\usepackage{color}
\usepackage{amsmath}
\usepackage{latexsym}
\usepackage[LGR,T1]{fontenc}
\usepackage[english]{babel}

\newcommand{\qsq}{\mbox{Q$^2$ }}

\newcommand{\gevsq}{\mbox{GeV$^2$ }}

\newcommand{\xBj}{x_{\rm Bj}}  
\newcommand{\GeVsq}{GeV$^2${}}
\newcommand{\PbFtwo}{PbF$_2${}}
\newcommand{\phigg}{\phi_{\gamma\gamma}}  
\newcommand{\Real}{\Re\text{e}}
\newcommand{\Imag}{\Im\text{m}}

\begin{document}

\title{Measurements of the Electron-Helicity Dependent Cross Sections
of Deeply Virtual Compton Scattering with CEBAF at 12 GeV}

\author{ {\bf Julie Roche}}
\altaffiliation{Co-Spokesperson}
\affiliation{Rutgers, The State University of New Jersey, Piscataway, New Jersey 08854; and\\
Thomas Jefferson National Accelerator Facility, Newport News, Virginia 23606}
\altaffiliation{Permanent Address: Ohio University, Athens OH, 45701}
\author{{\bf Charles E. Hyde-Wright}\footnote{Contact Person: chyde@odu.edu}}
\altaffiliation{Co-Spokesperson}
\author{ G. Gavalian, M. Amarian, S.~B\"ultmann,
G.E. Dodge, H. Juengst, J. Lachniet, A.~Radyushkin,  P.E. Ulmer, 
L.B. Weinstein}
\affiliation{Old Dominion University, Norfolk VA}

\author{{\bf Bernard Michel}}
\altaffiliation{Co-Spokesperson}
\author{J. Ball, P.-Y.~Bertin \footnote{And 
Thomas Jefferson National Accelerator Facility, Newport News, Virginia 23606}, 
M.~Brossard, R. De Masi, M. Gar\c con, F.-X. Girod, M. Guidal,
M.~Mac~Cormick, M. Mazouz,  S. Niccolai, B. Pire,
S. Procureur, F. Sabati\'e, E. Voutier, S. Wallon}
\affiliation{
LPC (Clermont) / LPSC (Grenoble) / IPNO \& LPT (Orsay) /
CPhT-Polytechnique (Palaiseau) / SPhN (Saclay)\\
\small CEA/DSM/DAPNIA \& CNRS/IN2P3, France}
\author{{\bf Carlos Mu\~ noz Camacho}}
\altaffiliation{Co-Spokesperson}
\affiliation{Los Alamos National Laboratory, Los Alamos NM, 87545}
\author{A. Camsonne, J.-P.~Chen, E. Chudakov, A. Deur, D.~Gaskell,
D.~Higinbotham, C.de Jager, J. LeRose, O. Hansen, R. Michaels,
S. Nanda, A.~Saha, S.~Stepanyan, B.~Wojtsekhowski
}
\affiliation{Thomas Jefferson National Accelerator Facility, Newport News, Virginia 23606}
\author{P.E.C.~Markowitz}
\affiliation{Florida International University, Miami FL}
\author{X. Zheng}
\affiliation{Massachusetts Institute of Technology, Cambridge MA 02139\,}
\altaffiliation{Permanent Address: University of Virginia, Charlottesville, VA 22904}
\author{R. Gilman, X. Jiang, E. Kuchina, R. Ransome}
\affiliation{Rutgers, The State University of New Jersey, Piscataway, NJ 08854}
\author{A. Deshpande}
\affiliation{Stony Brook University, Stony Brook, NY 11794}
\author{N. Liyanage}
\affiliation{University of Virginia, Charlottesville, VA 22904}
\author{Seonho Choi, Hyekoo Kang, Byungwuek Lee, Yumin Oh, Jongsog Song}
\affiliation{Seoul National University, Seoul 151-747, Korea}
\author{S. Sirca}
\affiliation{Dept. of Physics, University of Ljubljana, Slovenia}
\date{ 07 July 2006 for JLab PAC30}

\begin{abstract}
We propose precision measurements of the helicity-dependent and
helicity independent cross sections for the $ep\rightarrow ep\gamma$
reaction in Deeply
Virtual Compton Scattering (DVCS) kinematics.  DVCS scaling is obtained 
in the limits $Q^2\gg \Lambda_{\rm QCD}^2$,
$\xBj$ fixed, and $-\Delta^2 = -(q-q')^2\ll Q^2$.  We consider
the specific kinematic range $Q^2>2$ \GeVsq, $W>2$ GeV, and
$-\Delta^2\le1$ \GeVsq.  We will use our successful technique from
the 5.75 GeV Hall A DVCS experiment (E00-110).  With  polarized
 6.6, 8.8, and 11 GeV beams incident on the liquid hydrogen
 target, we will detect the scattered electron in
the Hall A HRS-L spectrometer (maximum central momentum 4.3 GeV/c)
and the emitted photon in a slightly expanded \PbFtwo{} calorimeter.
In general, we will not detect the recoil proton.
The H$(e,e'\gamma)X$ missing mass resolution is sufficient
to isolate the exclusive channel with $3\%$ systematic
precision.
\end{abstract}

\maketitle

\newpage
\tableofcontents
\newpage
\section{Introduction}

\subsection{Imaging the Nucleus}

We have a quantitative understanding of the strong interaction processes 
at extreme short distances in terms of perturbative QCD.  We also 
understand the long distance properties of hadronic interactions in terms 
of chiral perturbation theory.  At the intermediate scale: the scale of 
quark confinement and the creation of [ordinary] mass we have an 
understanding of numerous observables at about the 20\% level from lattice 
QCD calculations and semi-phenomenological models.  This is an extremely 
impressive intellectual achievement.  However, the questions we have today 
about nuclear physics, are questions at the 1\%, or even 0.1$\%$, level 
relative to the confinement scale $\Lambda_{\rm QCD}\approx 300$ MeV/c. 
For example, the $n$-$p$ mass splitting of 1.3 MeV and the Deuteron 
binding energy of 2.2 MeV are $\le 1\%$ effects that are crucial to the 
evolution of the universe.  The patterns of binding energies of neutron 
and proton rich nuclei are even smaller effects, and are crucial to the 
synthesis of elements $Z>$Fe in supernovae and other extreme events.  It 
is the QCD dynamics at the distance scale of $1/\Lambda_{\rm QCD}$ that 
gives rise to the origin of mass. To improve our understanding of 
confinement and of the origin of mass, we cannot rely solely on 
improvements in theory.  We must have experimental observables of the 
fundamental degrees of freedom of QCD--the quarks and gluons--at the 
distance scale of confinement. The generalized parton distributions (GPDs) 
are precisely the necessary observables.

Measurements of electro-weak form factors determine the spatial structure 
of charges and currents inside the nucleon. However, the resolution scale 
$Q^2$ is not independent of the distance scale $1/\sqrt{Q^2}$ probed. Deep 
inelastic scattering of leptons (DIS) and related inclusive high $p_\perp$ 
hadron scattering measure the distributions of quarks and gluons as a 
function of light cone momentum fractions, but integrated over spatial 
coordinates. Ji \cite{Ji:1996ek}, Radyushkin \cite{Radyushkin:1997ki}, and 
M\"uller {\it et al.\/} \cite{Mueller:1998fv}, defined a set of light cone 
matrix elements, now known as GPDs, which relate the spatial and momentum 
distributions of the partons. This allows the exciting possibility of 
determining spatial distributions of quarks and gluons in the nucleon as a 
function of their wavelength.

\begin{figure}
\begin{center}
\includegraphics[width=\linewidth]{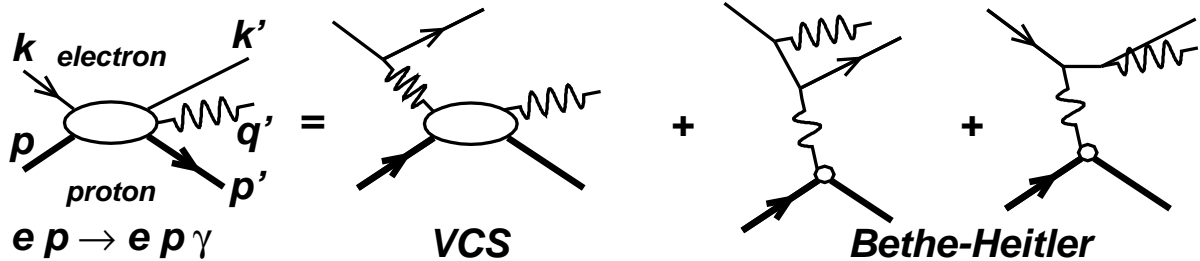}
\caption[{}VCS Feynman Diagram]{\label{fig:DVCSfeynman} Lowest order QED 
diagrams for the process $ep\rightarrow ep\gamma$, including the DVCS and 
Bethe-Heitler (BH) amplitudes. The external momentum four-vectors are 
defined on the diagram. The virtual photon momenta are $q^\mu=(k-k')^\mu$ 
in the DVCS- and $\Delta^\mu=(q-q')^\mu$ in the BH-amplitudes.
}
\end{center}
\end{figure}

\begin{figure}
\includegraphics[width=0.5\linewidth]{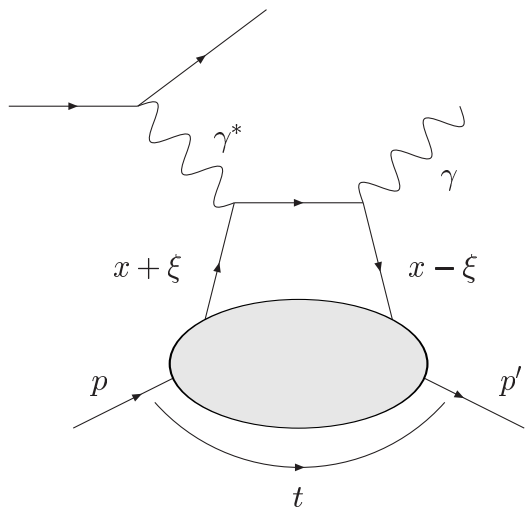}
\caption[DVCS Handbag Amplitude]{\label{fig:Handbag} Leading twist 
$\gamma^\ast p \rightarrow \gamma p$ amplitude in the DVCS limit. The 
initial and final quarks carry light-cone momentum fractions $x+\xi$ and 
$x-\xi$ of the light-cone momenta $(1\pm\xi)(p+p')^+$ (Eq. 
\ref{eq:lightcone}).  The crossed diagram is also included in the full 
DVCS amplitude. The invariant momentum transfer squared to the proton is 
$t=\Delta^2$.}
\end{figure}

The factorization proofs \cite{Ji:1998xh,Collins:1998be} established that 
the GPDs are experimentally accessible through deeply virtual Compton 
scattering (DVCS) and its interference with the Bethe-Heitler (BH) process 
(Fig.~\ref{fig:DVCSfeynman}). In addition, the spatial resolution ($Q^2$) 
of the reaction is independent of the distance scale $\approx 
1/\sqrt{t_{\rm min}-t}$ probed. Quark-Gluon operators are classified by 
their twist: the dimension minus spin of each operator.  The handbag 
amplitude of Fig.~\ref{fig:Handbag} is the lowest twist (twist-2) 
$\gamma^\ast p \rightarrow \gamma p$ operator. The factorization proofs 
confirm the connection between DIS and DVCS.  The proofs therefore suggest 
(but do not establish) that, just as in DIS, higher twist terms in DVCS 
will be only a small contribution to the cross sections at the $Q^2$ and 
$\xBj$ range accessible with electrons from 6--12 GeV.

In the formalism we are using \cite{Belitsky:2001ns}, the matrix elements 
of operators of twist-$n$ are $Q^2$ independent (except for 
$\ln(Q^2/\Lambda_{\rm QCD}^2)$ evolution), but all observables of 
twist-$n$ operators carry kinematic pre-factors that scale as $[(t_{\rm 
min}-t)/Q^2]^{n/2}$, $[(-t)/Q^2]^{n/2}$, or $[M^2/Q^2]^{n/2}$. Diehl {\it 
et al.,\/} \cite{Diehl:1997bu} showed that the twist-2 and twist-3 DVCS-BH 
interference terms could be independently extracted from the 
azimuthal-dependence ($\phigg$, \S\ref{sec:Kinematics}) of the helicity 
dependent cross sections. Burkardt \cite{Burkardt:2000za} showed that the 
$t$-dependence of the GPDs at $\xi = 0$ is Fourier conjugate to the 
transverse spatial distribution of quarks in the infinite momentum frame 
as a function of momentum fraction.  Ralston and Pire 
\cite{Ralston:2001xs} and Diehl \cite{Diehl:2002he} extended this 
interpretation to the general case of $\xi\ne 0$.  Belitsky {\it et 
al.,\/} \cite{Belitsky:2003nz} describe the general GPDs in terms of quark 
and gluon Wigner functions.

These elegant theoretical concepts have stimulated an intense experimental 
effort in DVCS.  The H1 \cite{Aktas:2005ty} \cite{Adloff:2001cn} and ZEUS 
\cite{Chekanov:2003ya} collaborations at HERA measured cross sections for 
$x_{\rm Bj}\approx 2\xi\approx 10^{-3}$.  These data are integrated over 
$\phigg$ and are therefore not sensitive to the BH$\cdot$DVCS interference 
terms. The CLAS \cite{Stepanyan:2001sm} and HERMES 
\cite{Airapetian:2001yk} collaborations measured relative beam helicity 
asymmetries. HERMES has also measured relative beam charge asymmetries 
\cite{Ellinghaus:2002bq,unknown:2006zr} and CLAS has measured longitudinal 
target relative asymmetries \cite{Chen:2006na}. The HERA and HERMES 
results integrate over final state inelasticities of $M_X^2\le 2.9\,{\rm 
GeV}^2$ (or greater). Relative asymmetries are a ratio of cross section 
differences divided by a cross section sum.  In general, these relative 
asymmetries contain BH$\cdot$DVCS interference and DVCS$^\dagger$DVCS 
terms in both the numerator and denominator (the beam charge asymmetry 
removes the DVCS$^\dagger$DVCS terms only from the numerator). Absolute 
cross section measurements are necessary to obtain all DVCS observables.

In Hall A, E00-110\cite{E00110} and E03-106\cite{E03106}, we measured 
absolute cross sections for H$(\vec{e},e'\gamma)$p and D$(e,e'\gamma)pn$ 
at $\xBj=0.36$. The following section describes the methods and results of 
E00-110. We propose to continue this program using the same experimental 
techniques, while taking advantage of the higher beam energy available 
with CEBAF at 12 GeV.

\subsection{Review of Hall A DVCS E00-110 Methods and Results
\label{sec:E00110}}

\begin{figure}
\centerline{
\includegraphics[width=0.5\linewidth]{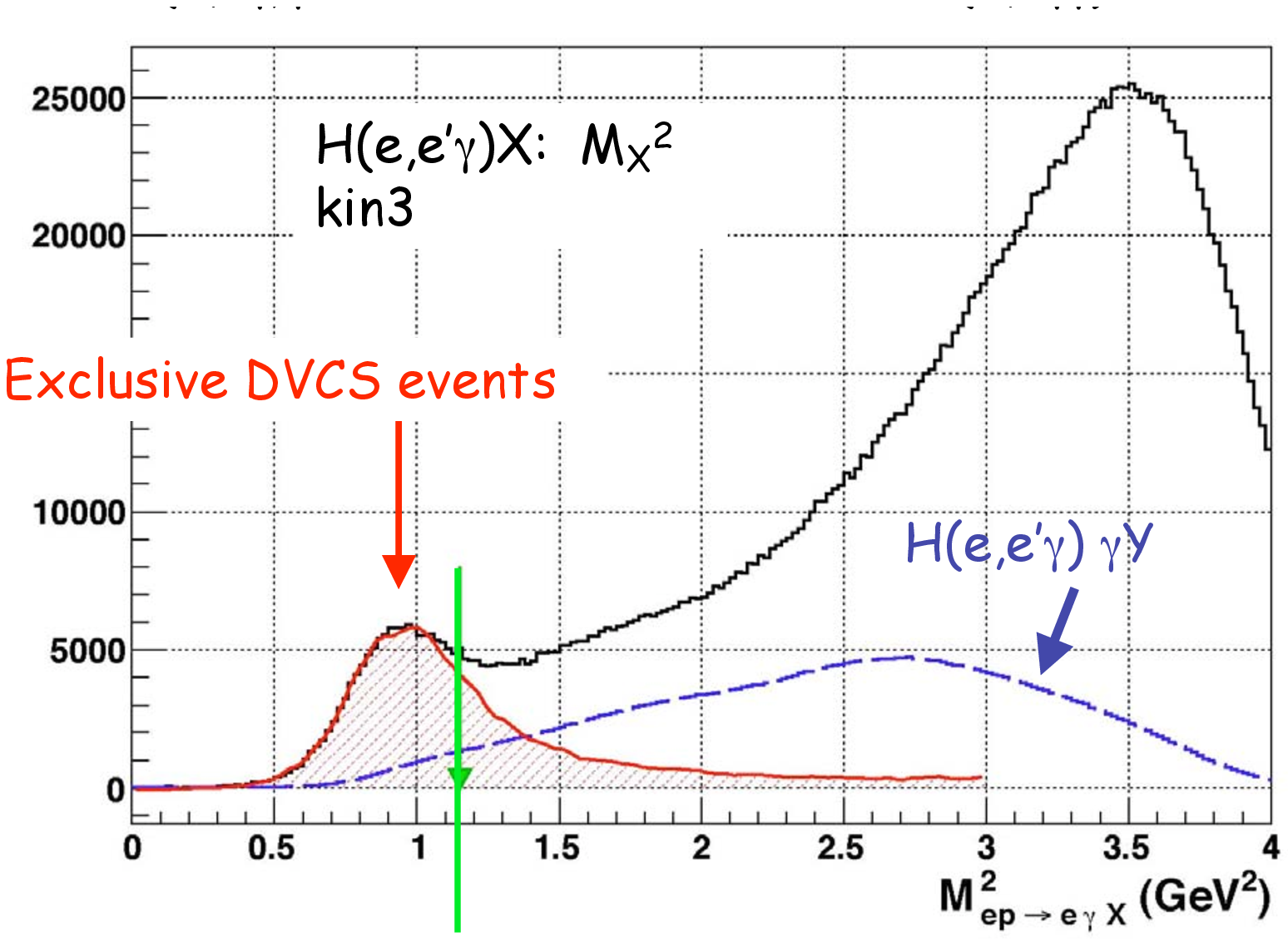}\hfil
\includegraphics[width=0.5\linewidth]{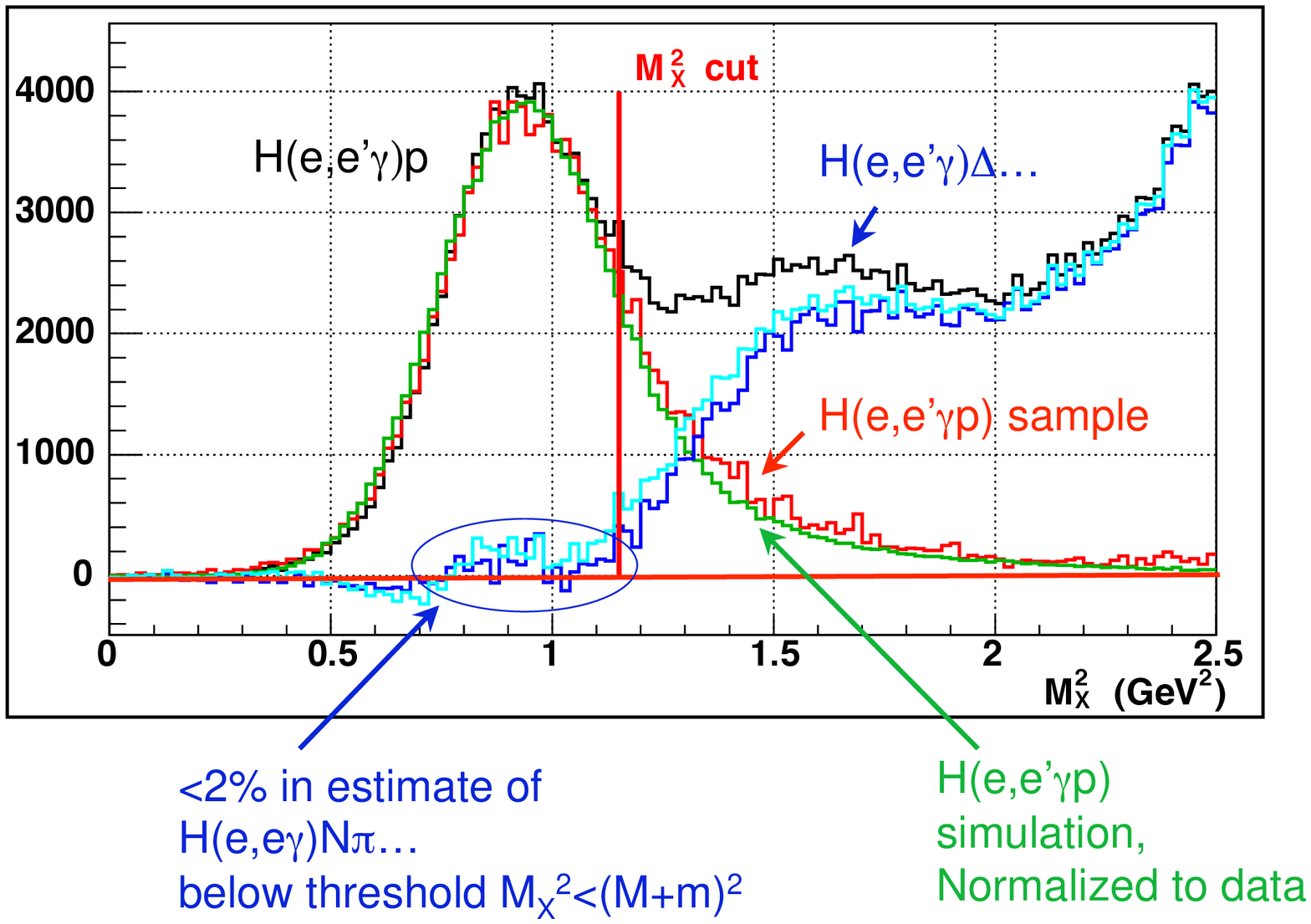}}
\caption{\label{fig:MX2E00110} Missing mass squared distribution for the 
H$(e,e'\gamma)X$ reaction, obtained in Hall A in E00-110. The left plot 
shows the raw spectrum (after accidental subtraction) and the statistical 
estimate of the contribution of $ep\rightarrow eY\pi^0$ events of the type 
H$(e,e'\gamma)\gamma Y$.  This estimate was obtained from the ensemble of 
detected H$(e,e'\pi^0)Y$ events, with both photons from the 
$\pi^0\rightarrow \gamma\gamma$ decay detected in the calorimeter. The 
right plot shows the H$(e,e'\gamma)X$ spectrum, after the $\pi^0$ 
subtraction.}
\end{figure}

The first draft publication of E00-110 can be found in 
Ref.~\cite{Camacho:2006hx}. Using a well understood experimental 
apparatus, this experiment measured both the unpolarized and polarized 
cross-section of the $\vec{e}p \rightarrow ep \gamma$ process in the 
Deeply Virtual Compton Scattering (DVCS) regime at $\xBj=0.36$ and for 
\qsq at 1.5, 1.9, and 2.3 \gevsq. With electrons detected in the HRS-L, we 
have an absolute acceptance for electron detection understood to 3\%, and 
a precise measurement of the scattering vertex and the direction of the 
virtual photon. DVCS is a three body final state, but at high $Q^2$ and 
low $t$, the final photon is highly aligned with the virtual photon and 
therefore highly correlated with the scattered electron. Thus, even with a 
modest calorimeter, our coincidence acceptance for DVCS is essentially 
limited only by the electron spectrometer.  As a consequence, very high 
values of the product of luminosity times coincidence acceptance are 
possible. The radiation hard \PbFtwo{} calorimeter gives a fast (Cerenkov) 
time response, and each channel is recorded with a 1 GHz digitizer which 
allows off-line identification of the DVCS photon. The identification of 
the exclusive channel is illustrated in Fig.~\ref{fig:MX2E00110}. In Hall 
A, our systematic errors are minimized by the combination of the Compton 
polarimeter, the well-understood optics and acceptance of the High 
Resolution Spectrometer (HRS), and a compact, hermetic, calorimeter. All 
of those factors allowed the measurements of cross-sections.  The 
high-precision electron detection minimizes systematic errors on $t$ and 
$\phi_{\gamma\gamma}$. Therefore, we exploit the precision 
$\phigg$-dependence to extract the cross section terms which have the form 
of a finite Fourier series modulated by the electron propagators of the BH 
amplitude.

The strong point of experiment E00-100 is that we measured absolute 
cross-sections. For example, the $\sin(\phi)$ term (modulated by BH 
propagators) of the helicity dependent cross section measures the 
interference of the imaginary part of the twist-2 DVCS amplitude with the 
BH amplitude, with a small contribution from an additional twist-3 
bilinear DVCS term.  The $Q^2$-dependence of this term places a tight 
limit on the contribution of the higher twist terms to our extraction of 
the ``handbag'' amplitude. The unpolarized cross section is a sum of the 
BH cross section, the real part of the BH$\cdot$DVCS interference, and a 
twist-2 bilinear DVCS term. The E00-110 results show that the unpolarized 
cross section is not entirely dominated by the BH cross section, but also 
has a large contribution from DVCS. Explicit twist-3 terms were also 
extracted from the $\sin(2\phigg)$ and $\cos(2\phigg)$ dependence of the 
cross sections.  These contribute very little to the cross sections.

In the kinematic regime of E00-100, neither the extracted Twist-2 or 
Twist-3 observables show any statistically significant dependence on \qsq. 
This provides strong support to the original theoretical predictions that 
DVCS scaling is based on the same foundation as DIS scaling.  We note that 
in the range $0.2\le\xBj\le0.6$, both the higher twist terms and $\ln 
Q^2/\Lambda_{QCD}^2$ evolution terms are small in DIS, even for 
$\qsq\approx 2$ GeV$^2$ 
\cite{Schaefer:2001uh,Osipenko:2004xg,Eidelman:2004wy}. Thus we are well 
on our way to proving the dominance of the leading twist term of the 
amplitudes ( ``handbag approximation'') where the virtual photon scatters 
off a single parton. This approximation is a corner stone of the study of 
the nucleon structure in terms of GPDs.

\subsection{Physics Goals and Proposed DVCS Kinematics 
\label{sec:PhysicsSummary}}

The present proposal cannot fully disentangle the spin-flavor structure of 
the GPDs.  We itemize here the measurements we will perform and the 
physical insights we expect to obtain.
\begin{itemize}
\item
Measure the $\vec{e} p \rightarrow e p \gamma$ cross sections at fixed 
$\xBj$ over as wide a range in $Q^2$ as possible for $k\le 11$ GeV.  This 
will determine with what precision the handbag amplitude dominates (or 
not) over the higher twist amplitudes.  More generally, we consider the 
virtual photon at high $Q^2$ as a superposition of a point-like elementary 
photon and a `hadronic' photon ($q\overline{q}$, vector mesons) with a 
typical hadronic transverse size.  The $Q^2$ dependence of the DVCS cross 
sections measures the relative importance of these two components of the 
photon \cite{Bauer:1977iq}.
\item Extract all kinematically independent observables (unpolarized 
target) for each $Q^2$, $\xBj$, $t$ point.  These observables are the 
angular harmonic superposition of Compton Form Factors (CFFs).  As 
functions of $\phigg$, the observable terms are $\cos(n\phigg)$ for 
$n\in\{0,1,2\}$, and $\sin(n\phigg)$ for $n\in\{1,2\}$, with additional 
$1/[J+K\cos(\phigg)]$ modulations from the electron propagators of the BH 
amplitude (\S \ref{sec:Belitsky}). Each of these five observables isolates 
the $\Real$ or $\Imag$ parts of a distinct combination of linear ($BH\cdot 
DVCS^\dagger$) and bilinear ($DVCS\cdot DVCS^\dagger$) terms.
\item Measure the $t$-dependence of each angular harmonic term. The 
$t$-dependence of each CFF is Fourier-conjugate to the spatial 
distribution of the corresponding superposition of quark distributions in 
the nucleon, as a function of quark light-cone momentum-fraction.  In a 
single experiment, we cannot access this Fourier transform directly, 
because we measure a superposition of terms.  However, we still expect to 
observe changes in the $t$-dependence of our observables as a function of 
$\xBj$. In particular, the $r.m.s.$ impact parameter of a quark of 
momentum fraction $x$ must diminish as a power of $(1-x)$ as $x\rightarrow 
1$. This is not a small effect, between $x=0.36$ and $x=0.6$, we expect a 
change in slope (as a function of $t$) of a factor two in individual GPDs.
\item Measure the $\vec{e} p \rightarrow e p \pi^0$ cross section in the 
same kinematics as DVCS. \\ This will test the factorization dominance of 
meson electro-production.  The longitudinal cross section ($d^2\sigma_L$) 
is the only leading twist (twist-2) term in the electro-production cross 
section. In this experiment, we do not propose Rosenbluth separations of 
$d^2\sigma_L$ from $d^2\sigma_T$.  However, as a function of $Q^2$, the 
ratio $d^2\sigma_T/d^2\sigma_L$ falls at least $\propto 1/Q^2$.  Thus the 
handbag contribution to $d^2\sigma_L$ can be extracted, within statistical 
errors, from a $1/Q^2$ expansion. The $\sigma_{LT}$, $\sigma_{TT}$, and 
$\sigma_{LT'}$ terms will be obtained from a Fourier decomposition of the 
azimuthal dependence of the cross section. These observables will provide 
additional constraints on both the longitudinal and transverse currents in 
pion electro-production.\\ The handbag amplitude of pion 
electro-production is the convolution of the axial GPDs $\tilde{H}$ and 
$\tilde{E}$ with the pion distribution amplitude (DA) $\Phi_\pi$.  For 
charged pion electro-production, the $\tilde{E}$ term is expected to be 
dominated by the pion-pole--this is the mechanism used to measure the pion 
form factor in electro-production on the proton.  However, neutral pion 
electro-production will be dominated by the non-pion-pole contributions to 
$\tilde{H}$ and $\tilde{E}$.
\end{itemize}

\begin{figure}
\includegraphics[width=\linewidth]{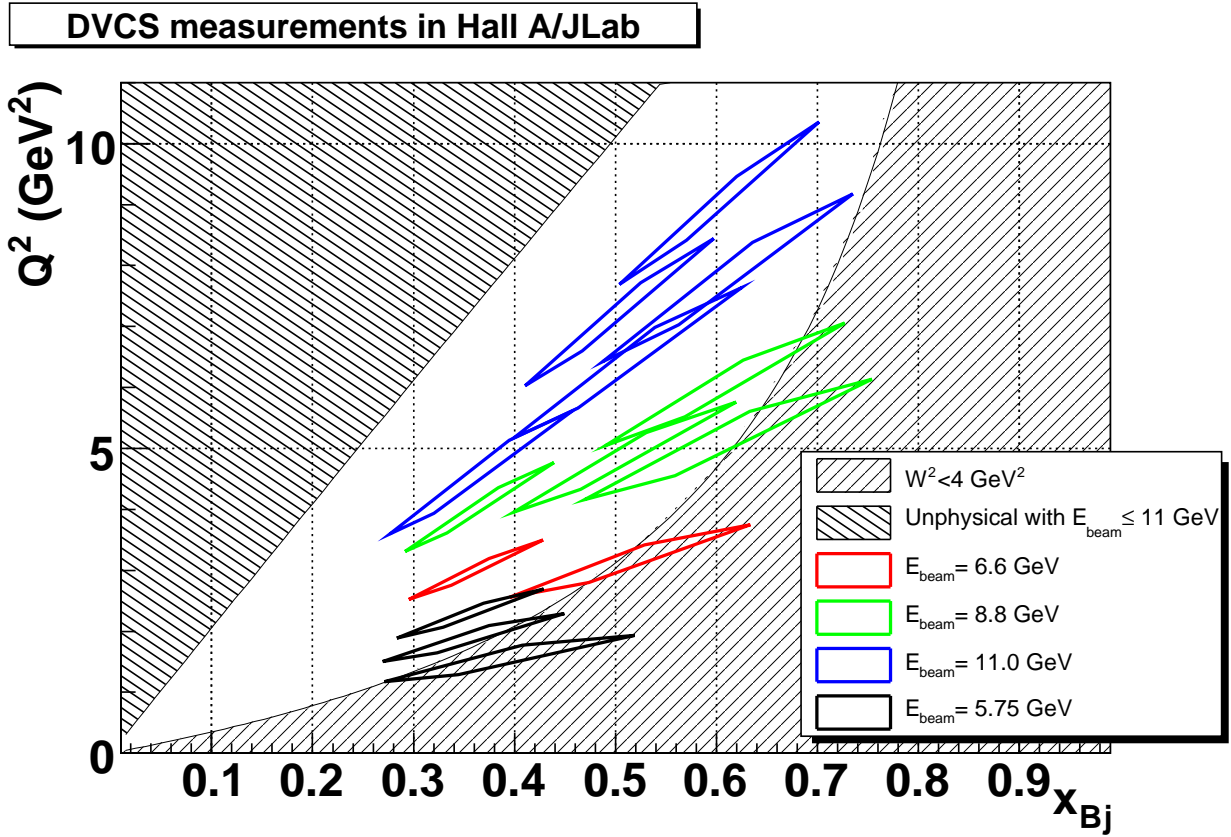}
\caption[Hall A DVCS Kinematics]{\label{fig:Q2xBj} Proposed DVCS 
kinematics for H$(e,e'\gamma)p$ measurements in Hall A with 3, 4, and 5 
pass beams of CEBAF at 12 GeV.  The diamond shapes trace the approximate 
acceptance of the HRS in each setting.  The boundary of the unphysical 
region corresponds to the maximum possible $Q^2$ at a given $\xBj$ for 11 
GeV.  This corresponds to $180^\circ$ electron scattering, equivalent to 
$\theta_q=0^\circ$. The points at $E_{\text Beam}=5.75$ GeV were obtained 
in E00-110.}
\end{figure}

To study the DVCS process with CEBAF at 12 GeV, we limit our definition of 
DVCS to the following kinematic range for the central values of our Hall A 
configurations:
\begin{eqnarray}
Q^2 &\ge& 2\,{\rm GeV}^2 \nonumber \nonumber\\
W^2 &\ge& 4\,{\rm GeV}^2 \nonumber \label{eq:DVCSlimits}\\
-t &<& Q^2
\end{eqnarray}
Our proposed kinematics, and the physics constraints of 
Eq.~\ref{eq:DVCSlimits} are illustrated in Fig.~\ref{fig:Q2xBj}. We 
propose a $Q^2$ scan of the cross sections for three values of $x_B$. Our 
maximum central $Q^2$ of 9 GeV$^2$ is higher than the $Q^2$ range of the 
$t$-distributions of published DVCS cross sections (integrated over 
$\phi_{\gamma\gamma}$) from HERA for $\xBj\approx 10^{-3}$. The beam time 
estimates for these high statistics measurements are listed in 
Table~\ref{tab:Q2xBjDays}.

Within the experimental constraints which we detail in section 
\ref{sec:Methods}, our choice of kinematics responds to the physics goals 
of high beam energy DVCS measurements.
\begin{itemize}
\item
At each $\xBj$ point, we measure the maximum possible range in $Q^2$. 
Although our preliminary data from E00-110 at $\xBj=0.36$ show no 
indication of $Q^2$ dependence in the observable $\Imag[{\mathcal 
C}^{\mathcal I}(\mathcal F)]$, this could result from a compensation of 
terms of higher twist and QCD evolution.  In this proposal, we will double 
the range in $Q^2$ at $\xBj=0.36$, and provide a nearly equal $Q^2$ range 
at $\xBj=0.5$ and $0.6$.
\item
Simple kinematics dictates that as the momentum fraction $x+\xi$ of the 
struck quark goes to 1, its impact parameter $b/(1+\xi)$ relative to the 
center of momentum of the initial proton must shrink towards zero. 
Burkardt has discussed the ``shrinkage'' of the proton as $x\rightarrow 1$ 
in the limit $\xi=0$. \cite{Burkardt:2004bv,Burkardt:2003mb}.  The 
perpendicular momentum transfer $\Delta_\perp$ is Fourier conjugate to the 
impact parameter $b$.  As a function of $x$, the mean square transverse 
separation between the struck quark and the center-of-momentum of the 
spectator system is
\begin{equation}
\langle{\bf r}_\perp^2\rangle(x) =\langle b^2\rangle/(1-x)^2,
\end{equation}
where $\langle b^2\rangle$ is the mean square impact parameter of the 
struck quark. Burkardt considered GPD models of the form \begin{equation} 
H_f(x,0,t)=q_f(x)\exp\left[at(1-x)^n/x\right]. \end{equation} The bound 
that $\langle{\bf r}_\perp^2\rangle$ remain finite as $x\rightarrow 1$ 
requires $n\ge 2$.  On the other hand, $n=1$ has been used extensively for 
modeling GPDs. At $x=0.6$, the choice of $n=1$ or $n=2$ changes the 
logarithmic slope $\partial \ln H/\partial t$ by a factor 2.5. This model 
illustrates that without measurements, there are very large uncertainties 
in the behavior of the GPDs at large $x$ (independent of the $t=0$ 
constraints), and that measurements will improve our understanding of the 
transverse distance scales of the quarks and gluons inside the proton.
\end{itemize}
Our specific choice of kinematics in Fig.~\ref{fig:Q2xBj} and 
Table~\ref{tab:Q2xBjDays} maximizes our range in $\xBj$, while maintaining 
measurements as a function of $Q^2$ for each value of $\xBj$.

\begin{table}
\begin{tabular}{cccccc} \hline \hline
$Q^2$& $k$     &\multicolumn{3}{c} {{\bf Beam Time}} &
{\bf Total} (Days)\\
(GeV$^2$)&(GeV)  & $\ x_{\rm Bj}=0.36\ $   & $\ \xBj=0.50\ $ & $\ \xBj=0.6\ $  &  (Days)    \\ \hline
3.0  & 6.6    &  3           &              &             &         \\
4.0  & 8.8    &  2           &              &             &         \\
4.55  &11.0   &  1           &              &             &         \\ \hline
3.1  & 6.6    &              &  5           &             &         \\
4.8  & 8.8    &              &  4           &             &         \\
6.3  &11.0    &              &  4           &             &         \\
7.2  &11.0    &              &  7           &             &         \\  \hline
5.1  & 8.8    &              &              &   13        &         \\
6.0  & 8.8    &              &              &   16        &         \\
7.7  &11.0    &              &              &   13        &         \\
9.0  &11.0    &              &              &   20        &         \\  \hline
TOTAL&        &   6          & 20           &   62        &  88     \\
\hline \hline
\end{tabular}
\caption[Beam Time Proposal]{Proposed Beam Time as a function of 
$(Q^2,k,\xBj)$ DVCS kinematics.}
\label{tab:Q2xBjDays}
\end{table}

\section{\label{sec:Observables} DVCS Observables}

\subsection{\label{sec:Kinematics}DVCS Kinematics and Definitions}

Fig.~\ref{fig:DVCSfeynman} defines our kinematic four-vectors for the $e p 
\rightarrow e p \gamma $ reaction.  The cross section is a function of the 
following invariants defined by the electron scattering kinematics:
\begin{eqnarray}
s_e &=& (k+p)^2 = M^2 + 2 k M \nonumber \\
Q^2 &=& - q^2 \nonumber \\
W^2 &=& (q+p)^2 = M^2 + Q^2\left[{1\over \xBj}-1\right] \nonumber \\
\xBj&=& {Q^2\over 2 q\cdot p} = {Q^2 \over W^2-M^2+Q^2} \nonumber \\
y &=& q\cdot p / k\cdot p.
\end{eqnarray}
The DVCS cross section also depends on variables specific to deeply 
virtual electro-production: the DVCS scaling variable $\xi$, the invariant 
momentum transfer squared $t$ to the proton,
azimuth $\phigg$ of the final photon around the ${\bf q}$-vector 
direction.  The DVCS scaling variable $\xi$ is defined in terms of the 
symmetrized momenta:
\begin{eqnarray}
\xi &=& {-\overline{q}^2 \over  \overline{q}\cdot P} \nonumber\\
 &=& \xBj { 1 + t/Q^2 \over 2-\xBj(1-t/Q^2) } \nonumber \\
 & & \rightarrow {\xBj \over 2 - \xBj}\ \ {\rm for}\ |t|/Q^2 <<1 \\
\overline{q}^\mu &=& (q+q')^\mu / 2 \nonumber \\
P^\mu  &=& (p+p')^\mu. \nonumber \\
P^2 &=& 4M^2 - t \nonumber \\
t &=& (p'-p)^2=\Delta^2.
\label{eq:xi}
\end{eqnarray}
In light cone coordinates $(p^+,{\bf p}_\perp,p^-)$, with $p^\pm = (p^0\pm 
p^z)/\sqrt{2}$, the four-vectors are:
\begin{eqnarray}
p &=& \left[(1+\xi)P^+,\frac{-\Delta_\perp}{2},\frac{M^2+\Delta_\perp^2/4}{2P^+(1+\xi)}\right] \nonumber \\
p' &=& \left[(1-\xi)P^+,\frac{+\Delta_\perp}{2},\frac{M^2+\Delta_\perp^2/4}{2P^+(1-\xi)}\right] \nonumber \\
P &=&  \left[P^+,0,\frac{4 M^2- t}{2P^+}\right]. \label{eq:lightcone}
\end{eqnarray}
In Fig.~\ref{fig:DVCSfeynman}, the initial (final) quark light cone 
momentum fraction is $x\pm\xi$ of the symmetrized momentum $P$ and 
$(x\pm\xi)/(1\pm\xi)$ of the initial (final) proton momentum $p^+$ 
($p^{\prime\,+}$).

Our convention for $\phigg$ is defined as the azimuthal angle in a 
event-by-event spherical-polar coordinate system (in the lab):

\begin{eqnarray}
\hat{z}_q &=& {\bf q} / |{\bf q}| \nonumber \\
\hat{y}_q &=& \left[{\bf k}\times {\bf k}'\right] / |{\bf k}\times {\bf k}'|
\nonumber \\
\hat{x}_q &=& \hat{y}_q \times \hat{z}_q. \\
\tan(\phigg) &=& \left. ({\bf q}'\cdot \hat{y}_q)\right/
                        ({\bf q}'\cdot\hat{x}_q)
\nonumber \\
       &=& \left.\left[ |{\bf q}| {\bf q}' \cdot({\bf k}\times{\bf k}' \right]\right/
                 \left[{\bf q}\times {\bf q}')\cdot({\bf k}\times{\bf k}' ) \right].
\label{eq:phigg}
\end{eqnarray}
The quadrant of $\phigg$ is defined by the signs of $({\bf q}'\cdot 
\hat{y}_q)$ and $({\bf q}'\cdot\hat{x}_q)$ We also utilize the laboratory 
angle $\theta_{\gamma\gamma}$ between the ${\bf q}$-vector and ${\bf 
q}'$-directions:
\begin{eqnarray}
\cos\theta_{\gamma\gamma} &=& \hat{z}_q \cdot {\bf q'}/|{\bf q'}|.
\end{eqnarray}
The impact parameter $b$ of the light-cone matrix element is Fourier 
conjugate to the $\Delta_\perp$, the momentum transfer perpendicular to 
the light cone direction defined by $P^+$:
\begin{eqnarray}
\Delta^2_\perp &=& (t-t_{\rm min})\frac{(1-\xi^2)}{(1+\xi^2)} \\
t_{\rm min}=t(\theta_{\gamma\gamma}=0^\circ) &=&
\frac{-4M^2\xi^2}{1-\xi^2} \approx \frac{- \xBj^2 M^2}{\left[
      1-\xBj(1-M^2/Q^2)\right]}.
\end{eqnarray}

\subsection{DVCS Cross Section\label{sec:Belitsky}}

The following equations reproduce the consistent expansion of the DVCS 
cross section to order twist-3 of Belitsky, M\"uller, and Kirchner 
\cite{Belitsky:2001ns}. Note that our definition of $\phi_{\gamma\gamma}$ 
agrees with the ``Trento-Convention'' for $\phi$ \cite{Bacchetta:2004jz}, 
and is the definition used in \cite{Airapetian:2001yk} and 
\cite{Stepanyan:2001sm}. Note also that this azimuth convention differs 
from $\phi_{\text{\cite{Belitsky:2001ns}}}$ defined in 
\cite{Belitsky:2001ns}{} by 
$\phi_{\gamma\gamma}=\pi-\phi_{\text{\cite{Belitsky:2001ns}}}$. In the 
following expressions, we utilize the differential phase space element 
$d^5\Phi=dQ^2 dx_{\rm Bj} d\phi_e dt d\phi_{\gamma\gamma}$. The 
helicity-dependent ($\lambda=\pm 1$) cross section for a lepton of charge 
$\pm e$ on an unpolarized target is:
\begin{eqnarray}
\frac{d^5\sigma(\lambda,\pm e)}{d^5\Phi} &=&
\frac{d\sigma_0}{dQ^2 d\xBj}
\left| \mathcal T^{BH}(\lambda) \pm \mathcal T^{DVCS}(\lambda)
     \right|^2/|e|^6 \nonumber \\
 &=&  \frac{d\sigma_0}{dQ^2 d\xBj}   \left[
       \left|\mathcal T^{BH}(\lambda) \right|^2 +
        \left| \mathcal T^{DVCS}(\lambda)\right|^2  \mp
        \mathcal I(\lambda)   \right]\frac{1}{e^6} \label{eq:dsigDVCS}
 \\
 \frac{d\sigma_0}{dQ^2 d\xBj} &=&
\frac{\alpha_{\rm QED}^3}{16\pi^2(s_e-M^2)^2 \xBj}
\frac{1}{\sqrt{1+\epsilon^2}}  \nonumber \\
\epsilon^2 &=& 4 M^2 \xBj^2/Q^2
\label{eq:dsig0}
\end{eqnarray}
The $|\mathcal T^{BH}|^2$ term is given in \cite{Belitsky:2001ns}, Eq.~25, 
and will not be reproduced here, except to note that it depends on 
bilinear combinations of the ordinary elastic form factors $F_1(t)$ and 
$F_2(t)$. The interference term $\mathcal I$ is:
\begin{eqnarray}
\frac{1}{e^6}\mathcal I &=&
\frac{1}{\xBj y^3 \mathcal P_1(\phigg) \mathcal P_2(\phigg) t }
\left\{ c_0^{\mathcal I} + \sum_{n=1}^3
        (-1)^n \left[     c_n^{\mathcal I}(\lambda)\cos(n\phigg)
        -    \lambda s_n^{\mathcal I}\sin(n\phigg) \right] \right\}.
\label{eq:IntPhi}
\end{eqnarray}
The $(-1)^n$ and $(-\lambda)$ factors are introduced by our convention for 
$\phigg$, relative to \cite{Belitsky:2001ns}. The bilinear DVCS terms have 
a similar form:
\begin{eqnarray}
\left| \mathcal T^{DVCS}(\lambda) \right|^2 \frac{1}{e^6} &=&
\frac{1}{y^2 Q^2} \left\{
c_0^{DVCS} + \sum_{n=1}^2 (-1)^n
c_n^{DVCS} \cos(n\phigg) + \lambda s_1^{DVCS} \sin(\phigg) \right\}
\label{eq:DVCSPhi}
\end{eqnarray}
The Fourier coefficients $c_n$ and $s_n$ will be defined below 
(Eq.~\ref{eq:ci}--\ref{eq:cdvcs}).  The $c_n^{\mathcal I}$ and 
$s_n^{\mathcal I}$ are linear in the GPDs, the $c_n^{\mathcal DVCS}$ and 
$s_n^{\mathcal DVCS}$ are bi-linear in the GPDs (and their higher twist 
extensions). All of the $\phigg$-dependence of the cross section is now 
explicit in Eq.~\ref{eq:IntPhi} and \ref{eq:DVCSPhi}.

The $\mathcal P_i(\phigg)$ are the electron propagators of the BH 
amplitude:
\begin{eqnarray}
Q^2  \mathcal P_1(\phigg) &=& (k-q')^2 \nonumber \\
Q^2  \mathcal P_2(\phigg) &=& (k'+q')^2.
\end{eqnarray}
After some algebra, the kinematic dependence of the pre-factor of 
Eq.~\ref{eq:IntPhi} is
more apparent:
\begin{eqnarray}
\frac{1}{\xBj y^3 \mathcal P_1(\phigg) \mathcal P_2(\phigg) t }
&=& \frac{-(1+\epsilon^2)^2(s_e-M^2)/t}{Q^2
\left(J-2K\cos\phigg\right)\left(1+J+(t/Q^2)-2K\cos\phigg\right)}
\label{eq:IntPhi0} \\
J &=& \left[1-y - \frac{y\epsilon^2}{2}\right] \left(1+\frac{t}{Q^2}\right)
       +\left(1-\xBj\right)\left(2-y\right)\left(\frac{-t}{ Q^2} \right)
\nonumber \\
& & \rightarrow \left[1-y-\frac{y\epsilon^2}{2}\right]
\ \ {\rm as}\ t/Q^2\rightarrow 0 \\
K^2 &=& \frac{t_{\rm min}-t}{Q^2} \left(1-\xBj\right)
         \left[1-y-\frac{y^2\epsilon^2}{4}\right]
\left[\sqrt{1+\epsilon^2} - \frac{\xBj W^2}{W^2-M^2}
   \frac{(t_{\rm min}-t)}{Q^2}\right] \nonumber \\
 & & \rightarrow \frac{t_{\rm min}-t}{Q^2} \left(1-\xBj\right)
         \left[1-y\right]
\ \ {\rm as}\ t/Q^2\rightarrow 0
\end{eqnarray}
The coefficient $K$ appears not only in the BH propagators, but also in 
the kinematic prefactors of the Fourier decomposition of the cross section 
(see below).  For fixed $Q^2$ and $\xBj$, $K$ depends on $t$ as $(t_{\rm 
min}-t)/Q^2$.

\subsection{Fourier Coefficients and Angular Harmonics}

The Fourier coefficients $c^{\mathcal I}_n$ and $s^{\mathcal I}_n$ of the 
interference terms are:
\begin{eqnarray}
c_0^{\mathcal I} &=&
-8(2-y)\Real\left\{\frac{(2-y)^2}{1-y}K^2  \mathcal C^{\mathcal I}(\mathcal F)
+ \frac{t}{Q^2}(1-y)(1-\xBj)
\left[ \mathcal C^{\mathcal I}+\Delta\mathcal C^{\mathcal I}\right](\mathcal F)
  \right\} \nonumber \\
\left\{{c_1^{\mathcal I} \atop \lambda s_1^{\mathcal I} }\right\} &=&
   -8K \left\{ {(2-2y+y^2) \atop - \lambda y (2-y) }\right\}
\left\{{ \Real \atop \Imag }\right\} \mathcal C^{\mathcal I}(\mathcal F)
\nonumber \\
\left\{{c_2^{\mathcal I} \atop \lambda s_2^{\mathcal I} }\right\} &=&
   \frac{-16K^2}{2-\xBj} \left\{ {(2-y) \atop - \lambda y  }\right\}
\left\{{ \Real \atop \Imag }\right\} \mathcal C^{\mathcal I}
(\mathcal F^{\rm eff})       \label{eq:ci}
\end{eqnarray}
The Fourier coefficients $c_3^{\mathcal I}$, $s_3^{\mathcal I}$ are gluon 
transversity terms.  We expect these to be very small in our kinematics, 
though it would be exciting if they generated a measureable signal. The 
${\mathcal C^I}$ and $\Delta{\mathcal C^I}$ amplitudes are the angular 
harmonic terms defined in Eqs. 69 and 72 of \cite{Belitsky:2001ns} (we 
have suppressed the subscript ``unp'' since our measurements are only with 
an unpolarized target). These angular harmonics depend on the interference 
of the BH amplitude with the set ${\mathcal F}= \{{\mathcal H,\,\mathcal 
E,\,\tilde{\mathcal H},\,\tilde{\mathcal E}}\}$ of twist-2 Compton form 
factors (CFFs) or the related set ${\mathcal F}^{\rm eff}$ of effective 
twist-3 CFFs:
\begin{eqnarray}
{\mathcal C^I}({\mathcal F}) &=& F_1(t){\mathcal H}(\xi,t)+
     \xi G_M(t) \tilde{\mathcal H}(\xi,t)
    -{t\over 4 M^2} F_2(t) {\mathcal E}(\xi,t)
\label{eq:gpds1} \\
{\mathcal C^I}({\mathcal F}^{\rm eff})
  &=& F_1(t){\mathcal H}^{\rm eff}(\xi,t)+
      \xi G_M(t) \tilde{\mathcal H}^{\rm eff}(\xi,t)
        -{t\over 4 M^2} F_2(t) {\mathcal E}^{\rm eff}(\xi,t)
\label{eq:gpds2}\\
\left[{\mathcal C^I}+\Delta{\mathcal C^I}\right]({\mathcal F}) &=&
           F_1(t){\mathcal H}(\xi,t) -{t\over 4 M^2} F_2(t) {\mathcal E}(\xi,t)
    - \xi^2 G_M(t) 
       \left[{\mathcal H}(\xi,t)+{\mathcal E}(\xi,t)\right].
\label{eq:gpds3}
\end{eqnarray}
Note that $\left[{\mathcal C^I}+\Delta{\mathcal C^I}\right]$ depends only 
on ${\mathcal H}$ and ${\mathcal E}$. The usual proton elastic form 
factors, $F_1$, $F_2$ and $G_M=F_1+F_2$ are defined to have negative 
arguments in the space-like regime. The Compton form factors are defined 
in terms of the vector GPDs $H$ and $E$, and the axial vector GPDs 
$\tilde{H}$ and $\tilde{E}$. For example ($f\in\{u,d,s\}$) 
\cite{Belitsky:2001ns}:
\begin{eqnarray}
{\mathcal H}(\xi,t) &=& \sum_{f}  \left[\frac{e_f}{e}\right]^2
\Biggl\{
         i\pi     \left[H_f(\xi,\xi,t) - H_f(-\xi,\xi,t)\right]
 \nonumber \\
  & & +
 {\mathcal P}
    \int_{-1}^{+1} dx
\left[ \frac{1}{\xi-x} - \frac{1}{\xi+x} \right]   H_f(x,\xi,t)\Biggr\}.
\label{eq:CFF}
\end{eqnarray}
Twist-3 CFFs contain Wandzura-Wilzcek terms, determined by the twist-2 
matrix elements, and dynamic $qGq$ twist-3 matrix elements. The twist-2 
and twist-3 CFFs are matrix elements of quark-gluon operators and are 
independent of $Q^2$ (up to logarithmic QCD evolution). The kinematic 
suppression of the twist-3 (and higher) terms is expressed in powers of 
$-t/Q^2$ and $(t_{\rm min}-t)/Q^2$ in {\it e.g.\/} the $K$-factor. This 
kinematic suppression is also a consequence of the fact that the the 
twist-3 terms couple to the longitudinal polarization of the virtual 
photon.

The bilinear DVCS Fourier coefficients are:
\begin{eqnarray}
c_0^{\text DVCS} &=&   2(2-2y+y^2)    \mathcal C^{\text DVCS}(\mathcal F,\mathcal F^\ast)
   \nonumber \\
\begin{Bmatrix}{c_1^{\text DVCS}}\\{ \lambda s_1^{\text DVCS} }\end{Bmatrix} &=&
\frac{8K}{2-\xBj}
\begin{Bmatrix}2-y\\-\lambda y\end{Bmatrix}
\begin{Bmatrix}\Real \\ \Imag \end{Bmatrix}
 \mathcal C^{\text DVCS}(\mathcal F^{\text eff},\mathcal F^\ast) 
    \label{eq:cDVCS}
\end{eqnarray}
The $c_2^{\text DVCS}$ coefficient is a gluon transversity term.

The DVCS angular harmonics  are
\begin{eqnarray}
\mathcal C^{DVCS}(\mathcal F,\mathcal F^\ast) &=& \frac{1}{(2-\xBj)^2} \left\{
4(1-\xBj)\left(\mathcal H\mathcal H^\ast
       + \tilde{\mathcal H}\tilde{\mathcal H}^\ast \right)-
\xBj^2 2\Real\left[ \mathcal H \mathcal E^\ast
      + \tilde{\mathcal H}\tilde{\mathcal E}^\ast \right]  \right.\nonumber \\
 & & \left. - \left( \xBj^2 + (2-\xBj)^2 \frac{t}{4M^2}\right)
       \mathcal E \mathcal E^\ast - \xBj^2\frac{t}{4M^2}
        \tilde{\mathcal E}\tilde{\mathcal E}^\ast \right\}.
\label{eq:cdvcs}
\end{eqnarray}
The twist-3 term $\mathcal C^{DVCS}(\mathcal F^{\rm eff},\mathcal F^\ast)$ 
has an identical form, with one CFF factor replaced with the set $\mathcal 
F^{\rm eff}$. The $\mathcal C^{DVCS}(\mathcal F_T,\mathcal F^\ast)$, 
appearing with a $\cos(2\phi_{\gamma\gamma})$ weighting, also has the same 
form as Eq.~\ref{eq:cdvcs}, but now with one set $\mathcal{F}$ replaced by 
the set $\mathcal{F}_T$ of (twist-2) gluon transversity Compton form 
factors.

\subsection{\label{sec:BHDVCS}BH$\cdot$DVCS Interference and Bilinear DVCS 
Terms}

The BH$\cdot$DVCS interference terms are not fully separable from the 
bilinear DVCS terms.  We analyze the cross section 
({\it e.g.}~\cite{Camacho:2006hx}) in the general form
\begin{eqnarray}
\frac{d^5\sigma}{d^5\Phi} &=& \frac{d^5\sigma^{BH}}{d^5\Phi}
  +\frac{1}{\mathcal P_1(\phigg) \mathcal P_2(\phigg)}\sum_n\left\{
K_{cn} \Real\left[\mathcal C_n^{\mathcal I,\rm exp} \right] \cos(n\phigg) \right. \nonumber \\
 & & \left.
+\lambda 
K_{sn} \Im\left[\mathcal C_n^{\mathcal I,\rm exp} \right] \sin(n\phigg) \right\}
\label{eq:IntExp}
\end{eqnarray}
The factors $K_{cn,sn}$ are the purely kinematic pre-factors defined in 
Eqs.~\ref{eq:dsigDVCS}--\ref{eq:ci}. The experimental coefficients 
$\Real,\Imag \mathcal C_n^{\mathcal I,\rm exp}$ include contributions from 
the bilinear DVCS terms, that mix into different orders in $\cos(n\phi)$ 
or $\sin(n\phi)$ due to the absence of the BH propagators $\mathcal P_1 
\mathcal P_2$ in the DVCS$^2$ cross section.

From Eq.~\ref{eq:IntPhi0}, \ref{eq:IntPhi}, and \ref{eq:DVCSPhi}, we 
obtain the generic enhancement of the interference terms over the 
DVCS$^\dagger$DVCS terms (of the same order in $\sin(n\phigg)$ or 
$\cos(n\phigg)$):
\begin{eqnarray}
\left|\frac{BH\cdot DVCS}{DVCS^\dagger DVCS}\right|
&\propto& \frac{y^2 (s_e-M^2)}{-t}.
\end{eqnarray}

In each setting $(\xBj,Q^2)$ setting, for each bin in $t$, we therefore 
have the following experimental Twist-2 DVCS observables
\begin{eqnarray}
\Imag[\mathcal C^{\mathcal I,\rm exp}(\mathcal F)] &=& 
\Imag[\mathcal C^{\mathcal I}(\mathcal F)]+\langle \eta_{s1}\rangle 
\Imag[\mathcal C^{DVCS}(\mathcal F^\ast,\mathcal F^{\rm eff}) ] 
\label{eq:ImC1exp}\\
\Real\left\{[\mathcal C+\Delta\mathcal C]^{\mathcal I,\rm exp}(\mathcal F)\right\}
&=&
\Real\left\{[\mathcal C^{\mathcal I}+\Delta\mathcal C^{\mathcal I}](\mathcal F)\right\}
+ \langle\eta_0\rangle\, 
  \Real\left[\mathcal C^{\rm DVCS}(\mathcal F^\ast,\mathcal F) \right] \\
\Real\left\{\mathcal C^{\mathcal I,\rm exp} (\mathcal F)\right\} 
&=&
\Real\left\{\mathcal C^{\mathcal I}(\mathcal F)\right\}
+ \langle\eta_{c1}\rangle\, 
  \Real\left\{\mathcal C^{\rm DVCS}(\mathcal F^\ast,\mathcal F)\right\}.
\end{eqnarray}
The coefficients $\langle \eta_\Lambda \rangle$ are the acceptance 
averaged ratios of the kinematic coefficients of the bilinear DVCS terms 
to the BH$\cdot$DVCS terms.
In addition, we have the experimental Twist-3 DVCS observables:
\begin{eqnarray}
\Imag[\mathcal C^{\mathcal I,\rm exp}(\mathcal F^{\rm eff})]
&=&
\Imag[\mathcal C^{\mathcal I}(\mathcal F^{\rm eff})]+\langle\eta_{s2}\rangle 
\Imag[\mathcal C^{DVCS}(\mathcal F^\ast,\mathcal F^{\rm eff}))] 
\label{eq:ImCIeff}\\
\Real[\mathcal C^{\mathcal I,\rm exp}(\mathcal F^{\rm eff})]
&=& \Real[\mathcal C^{\mathcal I}(\mathcal F^{\rm eff})]+\langle\eta_{c2}\rangle 
\Real[\mathcal C^{DVCS}(\mathcal F^\ast,\mathcal F^{\rm eff}) ].
\label{eq:ReCIeff}
\end{eqnarray}

The values of the $\eta_\Lambda$ coefficients in the E00-110 kinematics 
are summarized in Table~\ref{tab:Bilinear}.  They are small, though they 
grow with $|t|$. The bilinear term in Eq.~\ref{eq:ImC1exp} is a Twist-3 
observable, therefore the coefficient $\langle\eta_{s1}\rangle$ will 
decrease as $1/\sqrt{Q^2}$. Based on the values of 
Table~\ref{tab:Bilinear}, and using our results in E00-110 to estimate 
$\Imag[\mathcal C^{DVCS}(\mathcal F^\ast,\mathcal F^{\rm eff}))] $, we 
conclude that the bilinear term likely makes less than a 10\% contribution 
to $\Imag[\mathcal C^{\mathcal I \text exp}(\mathcal F)]$. In any case, 
any comparison of the experimental results with model calculations, or fit 
of model GPDs to the observables, must include the bilinear terms, with 
the experimental values of $\langle \eta_\Lambda\rangle$.

\begin{table}[hb]
\caption{\label{tab:Bilinear} Weighting factors of bilinear DVCS terms
for BH$\cdot$DVCS observables in E00-110.}
\begin{ruledtabular}
\begin{tabular}{cccccc} 
$t$ (GeV$^2$)  &$-0.37$&$-0.33$&$-0.27$&$-0.23$&$-0.17$\\ \hline
$\langle \eta_{s1}\rangle$
& -0.0142   & -0.0120   &  -0.0099   & -0.0080   &  -0.0060  \\ 
$\langle \eta_{s2}\rangle$
& -0.048 &-0.042&-0.036&-0.030&-0.023\\
$\langle \eta_{c1}\rangle$
&-0.050&-0.048&-0.038 &-0.033 &-0.026\\
$\langle \eta_{0}\rangle$
&+0.015&+0.024&+0.031& +0.039&+0.045\\
$\langle \eta_{c2}\rangle$
&-0.038& -0.030&-0.022& -0.014&  -0.010\\
\end{tabular}
\end{ruledtabular}
\end{table}

\section{Description of Experiment Apparatus\label{sec:Methods}}

\begin{figure}
\includegraphics[width=\linewidth]{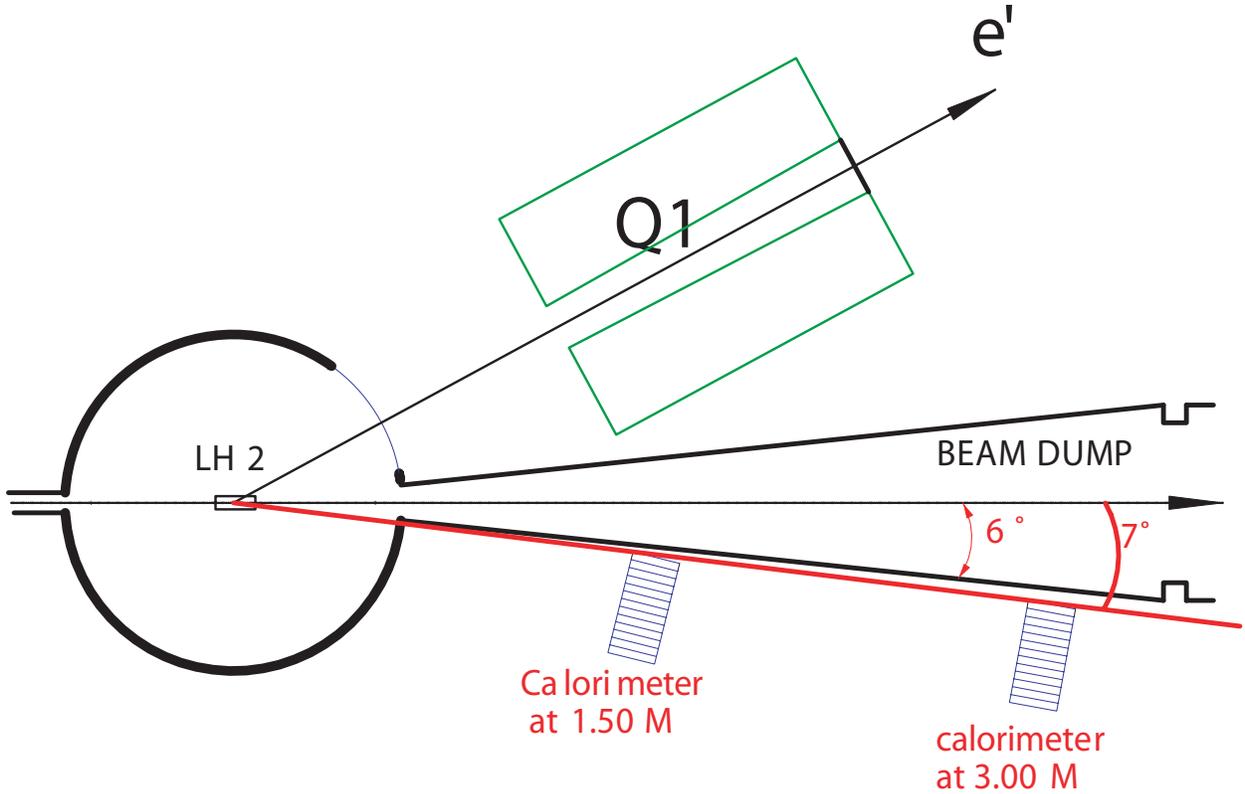}
\caption[Hall A DVCS Experimental Arrangement]{\label{fig:HallA} Hall A 
layout for DVCS with CEBAF at 12 GeV.  The scattering chamber is identical 
to the E00-110 chamber, with a midplane 63 cm radius spherical section 
with 1 cm Al wall thickness, and a 16 mil Al window facing the HRS L.  We 
propose a conical downstream beam pipe, with half-opening angle of 6 
degrees on beam-right, and length 3m. The drawing shows the expanded 
\PbFtwo{} calorimeter at beam-right, in the closest and farthest 
configurations: front face 150 cm and 300 cm, respectively, from target 
center. The calorimeter is shown in its smallest angle setting (inner edge 
at $7^\circ$).}
\end{figure}

\begin{table}
\caption{\label{tab:eekin} Detailed DVCS Kinematics.  The first line is 
from E00-110, and is included for comparison purposes.  The angle 
$\theta_q$ is the central angle of the virtual photon direction 
$q=(k-k')$.}
\begin{ruledtabular}\begin{tabular}{ccccccccc}
 $Q^2$ & $\xBj$ & $k$ & $k'$ & $\theta_e$   & $\theta_q$   & $q'(0^\circ)$ & $W^2$ \\
 (GeV$^2$) & & (GeV) & (GeV) & $({}^\circ)$ & $({}^\circ)$ & (GeV) &  (GeV$^2$) \\ \hline
  1.90 & 0.36 &   5.75 &  2.94 & 19.3 & 18.1 &  2.73 & 4.2 \\ \hline
  3.00 & 0.36 &   6.60 &  2.15 & 26.5 & 11.7 &  4.35 & 6.2  \\
  4.00 & 0.36 &   8.80 &  2.88 & 22.9 & 10.3 &  5.83 & 8.0 \\
  4.55 & 0.36 &  11.00 &  4.26 & 17.9 & 10.8 &  6.65 & 9.0 \\ \hline
  3.10 & 0.50 &   6.60 &  3.20 & 22.5 & 18.5 &  3.11 & 4.1 \\
  4.80 & 0.50 &   8.80 &  3.68 & 22.2 & 14.5 &  4.91 & 5.7 \\
  6.30 & 0.50 &  11.00 &  4.29 & 21.1 & 12.4 &  6.50 & 7.2 \\
  7.20 & 0.50 &  11.00 &  3.32 & 25.6 & 10.2 &  7.46 & 8.1 \\ \hline
  5.10 & 0.60 &   8.80 &  4.27 & 21.2 & 17.8 &  4.18 & 4.3 \\
  6.00 & 0.60 &   8.80 &  3.47 & 25.6 & 14.1 &  4.97 & 4.9 \\
  7.70 & 0.60 &  11.00 &  4.16 & 23.6 & 13.1 &  6.47 & 6.0 \\
  9.00 & 0.60 &  11.00 &  3.00 & 30.2 & 10.2 &  7.62 & 6.9 \\
\end{tabular}
\end{ruledtabular}
\end{table}

This proposal is based directly on the experience of E00-110. We present a 
sketch of the DVCS layout in Hall A in Fig.~\ref{fig:HallA}. We use the 
standard 15 cm liquid hydrogen target. We detect the electrons in the 
HRS-L and photons (and $\pi^0\rightarrow \gamma\gamma$) in a PbF$_2$ 
calorimeter at beam right. We note in Fig.~\ref{fig:HallA} the modified 
scattering chamber from E00-110 and a new modified downstream beam pipe.  
The scattering chamber is 63 cm in radius, with a 1 cm Al spherical wall 
facing the \PbFtwo{} calorimeter and a thin window (16 mil Al) facing the 
HRS-L.

The HRS-Left (HRS-L) limits the central values of the scattered electron 
momentum to
\begin{eqnarray}
k'&\le& 4.3 GeV/c, \nonumber \\
\theta_e &\ge& 12.5^\circ. \label{eq:kHRS}
\end{eqnarray}
As detailed in section \ref{sec:Calo}, to handle both the instantaneous 
pile-up and integrated radiation dose in the calorimeter, we limit the 
placement of the calorimeter to
\begin{equation}
\theta_{\rm PbF_2}^{\rm min}  \ge 7^\circ. \label{eq:CaloThetaMin}
\end{equation}
At the same time, to ensure adequate azimuthal coverage in 
$\phi_{\gamma\gamma}$, we limit the placement of the calorimeter
\begin{equation}
\theta_q \ge 10^\circ. \label{eq:ThetaqMin}
\end{equation}

The following subsections detail our technical solutions, and demonstrate 
that these technical constraints do not limit the physics scope of this 
proposal.

\begin{figure}
\includegraphics[width=\linewidth]{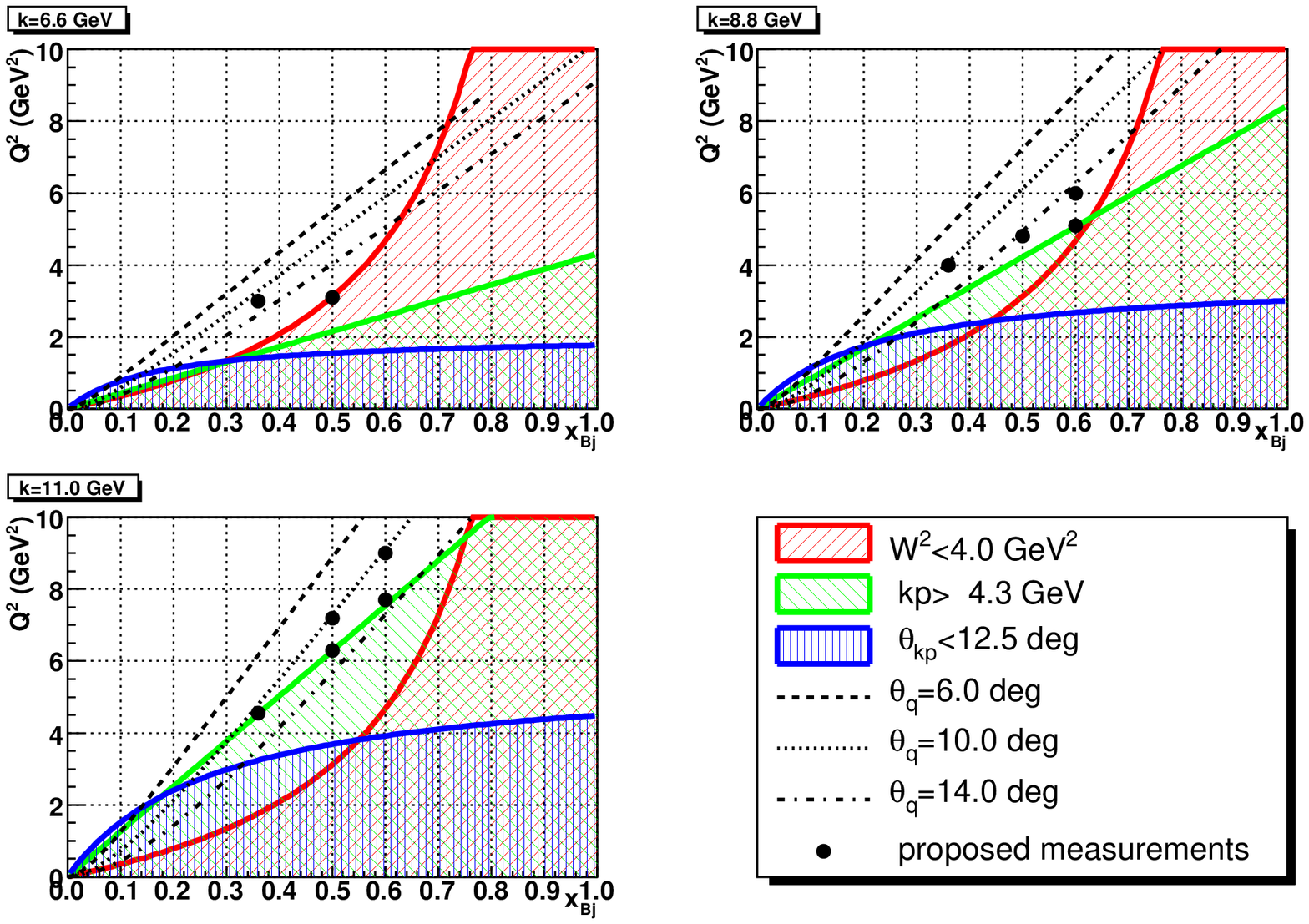}
\caption[DVCS Kinematics at 6.6, 8.8, \& 11 GeV]{\label{fig:DVCSkinEbeam} 
Proposed DVCS central kinematics for H$(e,e'\gamma)p$ measurements in Hall 
A with 6.6, 8.8, and 11 GeV incident beam.  The experimental constraints 
of Eq.~\ref{eq:kHRS},\ref{eq:ThetaqMin} are indicated.  Each experimental 
point is the center of one of the 'diamond' regions of 
Fig.~\ref{fig:Q2xBj}. }
\end{figure}

\subsection{High Resolution Spectrometer}

Our proposed kinematics, and the physics constraints of 
Eq.~\ref{eq:DVCSlimits} are illustrated in Fig.~\ref{fig:Q2xBj}. The 
individual beam energies are illustrated in Fig.~\ref{fig:DVCSkinEbeam}, 
with the HRS constraints (Eq.~\ref{eq:kHRS}) superimposed. We note the 
following points with regard to the HRS and calorimeter 
(Eq.~\ref{eq:ThetaqMin}) constraints:
\begin{itemize}
\item  The HRS constraints have no effect for $Q^2>2$ GeV$^2$ at $k=6.6$ 
GeV.
\item
At $k=8.8$ and 11 GeV, primarily the $k<4.3$ GeV constraint removes 
roughly the { lower} half of the $Q^2$ range at each $\xBj$.  This $Q^2$ 
range for each $\xBj$ is covered by the lower beam energies.
\item
According to the Base Equipment plan for CEBAF, the three Halls A, B, C, 
when running simultaneously, must operate at different multiples of 2.2 
GeV.
\item
The small angle calorimeter cut of $\theta_q>10^\circ$ removes the very 
highest physically allowed $Q^2$ values at each $\xBj$.
\item
All of the constraints prevent us from kinematics for $\xBj\le 0.2$.
\item
Kinematics at $\xBj=0.7$ are allowed at $k=8.8$ and 11 GeV.  At this time, 
it is very difficult to make reliable estimates of the DVCS signal at this 
extreme $\xBj$, and we do not include these kinematics in our proposal.
\end{itemize}
The detailed kinematics are summarized in Table~\ref{tab:eekin}.

The HRS performance is central to this experiment. Once we make a 
selection of exclusive H$(e,e'\gamma)p$ events, the resolution in 
$t=(q-q')^2$ is determined by approximately equal contributions from the 
HRS momentum resolution and the angular precision of the direction 
${\bf\hat{q}'}$.  The resolution on the direction of ${\bf q}'$ has equal 
contributions from the position resolution in the calorimeter and the 
vertex resolution, as obtained from the spectrometer.  The combination of 
precise acceptance, kinematic, and vertex resolution in the HRS makes it 
possible to make precision extractions of the DVCS observables.

\subsection{Beam Line}

In E00-110 and E03-106, with the front face of the calorimeter 1.10 m from 
target center, we had a 6'' diameter cylindrical beam pipe, welded 
directly to the scattering chamber.  This aperture corresponds to roughly 
a $6^\circ$ half-opening angle from target center. We require placing the 
calorimeter at distances from 1.5 m to 3.0 m from the target 
($\S$\ref{sec:Calo}). To avoid excess calorimeter background generated by 
secondaries striking the beam pipe, we propose a thin walled (1/8'' Al) 
conical beam pipe, welded to the same aperture.  The cone must be slightly 
off-axis, to prevent an interference with HRS-Q1 at the minimum 
spectrometer angle setting of $18^\circ$ (Table~\ref{tab:eekin}), while at 
the same time preserving a full $6^\circ$ aperture on Beam-right. The cone 
should be 3.0 m long, to continue downstream of the calorimeter. This will 
require moving or replacing a gate valve and pumping station.

\subsection{\label{sec:Calo}\PbFtwo{} Calorimeter}

\begin{table}[ht]
\caption[Properties of \PbFtwo.]{\label{tab:PbF2} Properties of \PbFtwo.}
\begin{ruledtabular}
\begin{tabular}{lll}
Density &  &7.77 g/cm$^3$ \\
Radiation Length & &0.93 cm \\
Moli\`ere Radius   & &2.20 cm \\
Index of Refraction & ($\lambda =180$ nm) & 2.05 \ \ \\ 
                    & ($\lambda= 400$ nm) & 1.82 \ \ \\
Critical Energy   &  & 9.04 MeV 
\end{tabular}
\end{ruledtabular}
\end{table}

We will detect the scattered photon in a $13\times 16$ element PbF$_2$ 
calorimeter. This is the existing $11\times 12$ E00-110 calorimeter, with 
76 additional elements. Each block is $3\times3$ cm$^2\times 20 X_0$. The 
additional blocks will add two more rows on the top and bottom, and two 
columns on the wide angle side. The properties of \PbFtwo{} are summarized 
in Table~\ref{tab:PbF2}. The important design considerations for DVCS are 
as follows.
\begin{itemize}
\item \PbFtwo{} is a radiation hard pure {C}erenkov crystal medium 
\cite{Woody:1994};
\item
With no scintillation light\cite{Anderson:1993ja}, the calorimeter signal 
is insensitive to low energy nuclear particles, and the pulse rise and 
fall time is determined only by geometry and the response of the PMT.  
This allows us to use the 1GHz Analog Ring Sampler (ARS) 
digitizer\cite{ars} to minimize pileup.
\item The high luminosity of this proposal requires fast response PMTs 
operated at low gain and capacitively coupled to a pre-amplifier.  The low 
gain reduces the DC anode current. The capacitive coupling removes the 
average pile-up from low energy $\gamma$-rays.
\item The small Moli\`ere radius (2.2 cm) allows us to separate closely 
space showers from $\pi^0$ decay, and minimize shower leakage at the 
boundary. \item The short radiation length minimizes fluctuations in light 
collection from fluctuations in the longitudinal profile of the shower.

\item The low value $9$ MeV of the critical energy (roughly the energy 
threshold for which bremsstrahlung energy loss exceeds ionization loss for 
electrons) also improves the resolution {\it e.g.} relative to Pb-Glass.
\item In E00-110, we obtained a signal of 1 photo-electron per MeV of 
deposited energy in the E.M. shower, and a energy resolution of 2.4\% from 
elastic H$(e,e'_{\text Calo}p_{\text HRS})$ electron of 4.2 GeV.  For our 
simulations, we project a resolution of $\sigma_E/E = 2.0\% \oplus 
(3.2\%)\sqrt{(1\,{\rm GeV})/q'}$.  We also achieved a spatial resolution 
of 2 mm at 4.2 GeV.  From the combination of energy and spatial 
resolution, we obtained a $\pi^0\rightarrow \gamma\gamma$ mass resolution 
of 9 MeV.
\end{itemize}
The size, granularity, and position of the calorimeter must accommodate 
the following constraints:
\begin{enumerate}
\item Nearly $2\pi$ azimuthal photon acceptance for $|\Delta_\perp|<0.6$ 
GeV/c. independent of the central kinematics.  The angular size required 
therefore shrinks as $\Delta_\perp^{\rm 
max}/q'(0^\circ)$.\label{item:acceptance}
\item Good separation of the two clusters from the $\pi^0\rightarrow 
\gamma \gamma $ decay, in order to measure the H$(e,e'\pi^0)p$ reaction. 
For $\pi^0\rightarrow \gamma\gamma$ reconstruction, we require a center to 
center separation of 3 \PbFtwo{} blocks, or 9 cm. In high energy DVCS or 
deep virtual $\pi^0$ kinematics, $q'_{\rm DVCS}\approx E_\pi$.  In this 
limit, the minimum half opening angle of the $\pi^0 \rightarrow 
\gamma\gamma$ decay is $\theta_{\pi\gamma} \ge m_\pi/q'$.  The minimum 
distance of the calorimeter from the target, based on the cluster 
separation requirement is given as the $D$ parameter in 
Table~\ref{tab:Q2xBjDays}. Furthermore, 50\% of the $\pi^0$ decay events 
yield both photons within a half angle cone of $\theta_{\pi\gamma} \le 
(\sqrt{3/2})(m_\pi/q')$.
\label{item:pion}
\item
Maximum distance from target to minimize pile-up and radiation dose per 
block at fixed luminosity.\label{item:lumi}
\end{enumerate}
Item \ref{item:pion} requires us to increase $D$ in proportion to $q'$. 
However, from item \ref{item:acceptance} we see that the acceptance is 
$\Delta_\perp$ remains invariant.  The solid angle per \PbFtwo{} block at 
the distance $D$ determines the maximum feasible luminosity for each 
setting. Therefore item \ref{item:lumi} allows us to increase the 
luminosity in proportion to $D^2\propto q^{\prime\,2}$.

We will calibrate the central blocks of the calorimeter via elastic 
H$(e,e'_{\text Calo}p_{\text HRS})$ measurements We anticipate three 
sequences of elastic measurements of one day each at the beginning, 
middle, and end of each scheduled period of running time. We will then 
cross calibrate all of the blocks and maintain a continuous monitor of the 
calibration with the $\pi^0$ mass reconstruction from H$(e,e'\pi^0)X$ 
events.  We anticipate sufficient statistics to obtain an independent 
calibration from each day of running. The elastic calibrations also serve 
to verify the geometrical surveys of the spectrometer and calorimeter.

\subsection{\label{sec:MaxLumi} Luminosity Limits and Optical Curing of 
Calorimeter}

In E03-106, we took $D(e,e'\gamma)X$ data at a maximum luminosity (per 
nucleon) of $4\cdot 10^{37}/$cm$^2$/s, with the front face of the PbF$_2$ 
calorimeter at 110 cm from the target center.  We did not obtain any 
degradation of resolution in the calorimeter from pileup of signals.  
During the entire 80 day run of E00-110 and E03-106, we delivered a total 
of 12 C to the 15 cm liquid Hydrogen and Deuterium targets.  We performed 
absolute calibrations of the calorimeter with elastic H$(e,e_{\rm 
Calo}^\prime p_{\rm HRS})$ events at the beginning and end of data taking.  
We observed up to 20\% decrease in signal amplitude in individual blocks, 
without observable loss in missing mass resolution after recalibration. 
Custom pre-amplifiers were used to keep the PMT anode current small. 
Therefore the loss of amplitude is attributed to degradation of the 
transmission properties of the blocks, and not to degradation of the 
photo-cathodes of the PMTs.  Independent numerical\cite{Pavel:2006ch} and 
analytic simulations indicate that below 10 degrees, the radiation dose to 
the calorimeter is dominated by M\o ller electrons (and related 
bremsstrahlung). Beyond 20 degrees, the dominant background arises from 
decay photons from inclusive $\pi^0$ photo-production. The simulations 
also indicate that from 7.5 to 11.5 degrees, the radiation dose diminishes 
by a factor 5.  On this basis, we conclude that roughly $50\%$ of the 
radiation dose received by the small angle blocks of the calorimeter 
occurred when the small angle edge of the calorimeter was at $7^\circ$.  
This is the same minimum angle we will use in the present proposal.

Achenbach {\it et al.\/} studied the radiation damage and optical curing 
of PbF$_2$ crystals\cite{Achenbach:1998yv}.  They found the radiation 
damage to be linear for doses up to 8 KGy (from a $^{60}$Co source).  For 
a dose of 1 KGy, they observed a loss of $25\%$ in transmission for blue 
light of $\lambda=400 $ nm.  They also obtained good results for curing 
the radiation damage by exposure to blue light. The front face of the 1000 
element A4 array was exposed for 17 hours to a Hg(Ar) pencil lamp 
(filtered to pass only $\lambda> 365 $ nm) at a distance of 50 cm 
(intensity on the calorimeter surface of $2\,\mu$W/cm$^2$). The 
transmission of 400 nm and 330 nm light returned to 100\% and 97\%, 
respectively, of its initial value.

At this stage, it is not possible to compare the absolute dose in our 
simulations with the dose recorded in the MAMI-A4 trials.  The A4 dose is 
recorded as a volumetric dose (1Gy = 1Joule/kg) yet the gamma rays from 
$^{60}$Co are predominantly absorbed in a layer of thickness 1/10 the 
transverse size of the crystals. The radiation dose in Hall A during a 
DVCS experiment is primarily from photons and electrons 50--1000 MeV.  
Thus we consider the absolute scale of dose comparison to be uncertain by 
a factor of 10. Instead, we normalize the future radiation damage of the 
calorimeter to the maximum value for the fractional signal attenuation per 
integrated luminosity obtained in conditions during E00-110 and E03-106 
that were nearly identical our proposed configuration. We utilize the 
studies from MAMI-A4 to determine that a radiation-dose induced 
attenuation of up to 25\% ($\lambda=400$ nm) can be cured (to within 
$1\%$) with a 17 hour exposure to blue light.  This curing must be 
followed by several dark hours to allow the phosphorescence of the PMT 
photo-cathodes to decay.

Pile-up within the 20 ns analysis window of the pulse shape analysis of 
the PbF$_2$ signals will limit our instantaneous luminosity. Based on our 
previous experience we can operate at an instantaneous luminosity times 
acceptance per PbF$_2$ block of
\begin{eqnarray}
\mathcal L(D) &=& \left[\frac{(4.0\cdot 10^{37})}{{\rm cm}^2\cdot {\rm s}}\right]
               \, \left[\frac{D}{(110\,{\rm cm})}\right]^2.
\label{eq:MaxLumi}
\end{eqnarray}
with $D$ the distance from target to calorimeter.  Our projected count 
rates and beam times are based on this luminosity as a function of the 
kinematic setting.

We plan to use blue light curing of the blocks every time the signal 
attenuation reaches $20\%$.  For those settings with the minimum edge of 
the calorimeter $\theta_{\rm Calo}^{\rm min}$ equal to $7^\circ$, this 
corresponds to 5 days of running at the luminosity of 
Eq.~\ref{eq:MaxLumi}.  We will use the $\pi^0\rightarrow \gamma\gamma$ 
mass resolution from both single arm H$(e,\pi^0)Y$ (prescaled) and 
coincidence H$(e,e'\pi^0)X$ events to monitor the light yields in the 
PbF$_2$ array. At larger calorimeter angle settings, the time between 
curing will be correspondingly longer.  We estimate a total of 10--12 
curing days during the experiment (in addition to the running time).

\subsection{Trigger}

The electron detector stack in the HRS will be the standard configuration 
of VDC I and II, segmented S1 and S2, gas Cerenkov, and Pb-Glass 
calorimeter.  The Cerenkov ($\check{C}er$) and Pb-Glass provide redundant 
electron identification in the off-line analysis.  The main HRS trigger is
\begin{eqnarray}
{\rm HRS} &=& \left[\cup_{i,j} (S1R_i\cap S1L_i)\cap(S2R_j\cap S2L_j) \right]
               \cap \check{C}er.
\end{eqnarray}
This requires a coincidence between the PMTs at the two ends of at least 
one scintillator paddle in each of $S1$ and $S2$.  In addition, we require 
a coincidence with the Cerenkov counter. Supplementary prescaled triggers 
with either $S1$, $S2$, or $\check{C}er$ removed from the coincidence will 
monitor the efficiency of each trigger detector.

In addition to the HRS electron trigger, we will upgrade our present 
coincidence validation/fast-clear logic. The calorimeter signals are 
continuously recorded by the 128 sample $\otimes$ 1GHz ARS 
\cite{ars,E00110} array. This array is digitized, at a slower rate, 
following a trigger validation. For each HRS trigger, we stop the ARS 
sampling, and trigger a Sample and Hold (SH) circuit, that is coupled to 
the calorimeter signals via a high impedance input (in the future, we may 
replace this with a pipeline ADC).  The SH signals are digitized and then 
in a field programmable gate array (FPGA), we will form the following 
trigger validation signals.
\begin{enumerate}
\item
A validation of the HRS-electron trigger, based on detecting at least one 
$2\times2$ cluster above a programmable threshold $E_\gamma^{\rm Th}$.
\item
A $\pi^0$ trigger based on detecting at least two separated $2\times2$ 
clusters, with each cluster above a programmable threshold $E_1$ and the 
sum of the two clusters above a programmable threshold $E_\pi$.  This 
trigger will select candidate H$(e,e'\pi^0)X$ events.
\end{enumerate}
If a valid signal is found, the ARS array is digitized and recorded 
(together with the HRS signals).  If the HRS trigger is not validated by 
the calorimeter signals, a fast clear is issued to the ARS array, no 
digitization occurs, and acquisition resumes.  During E00-110, the 
SH/Fast-clear cycle took 500 ns.  With upgrades to the FPGA, we can 
shorten this deadtime by a factor of two.  In addition, upgrades to the 
VME standard will allow us to increase the total bandwidth of data 
acquisition for this proposal.

\section{Projected Results\label{sec:Results}}

\begin{table}
\begin{ruledtabular}
\caption[Experimental Conditions]{\label{tab:Trues} Experimental 
Conditions for DVCS. For each $(Q^2,k,\xBj)$ setting, we present: Maximum 
photon energy $q'(0^\circ)$; Calorimeter distance $D$; Virtual photon 
direction $\theta_q$; Angle $\theta_{calo}^{min}$ of the edge of the 
calorimeter, relative to the beam line; Kinematic minimum $|t_{\rm min}|$ 
and upper bound $|t_{\rm max}|$ for approximately full acceptance in 
$\phigg$; Resolution $\sigma(M_X^2)$ in H$(e,e'\gamma)X$ missing mass 
squared; Luminosity $\mathcal L$; H$(e,e)X$ inclusive trigger rate, 
H$(e,e'\gamma)p$ exclusive DVCS count rate. The beam time is calculated to 
obtain an estimated 250K events at each setting, or at least 40,000 events 
per bin in $t_{\rm min}-t$.  The distance $D$ of the front face of the 
calorimeter from the target center is optimized for the separation of the 
clusters from $\pi^0\rightarrow \gamma\gamma$ decay 
(section~\ref{sec:Calo}). The intrinsic missing mass resolution in E00-110 
is $\sigma(M_X^2) = 0.20$ GeV$^2$. The luminosity is determined by the 
maximum rate allowed by pileup in the calorimeter.  This luminosity is 
proportional to $D^2$ ($\S$\ref{sec:MaxLumi}). }
\begin{tabular}{|ccc|cc|cc|cc|c|c|cc|c|}
$Q^2$& $k$    & $\xBj$ & $ q'(0^\circ)$ & $D$   & $\theta_q$  &$\theta_{calo}^{min}$  & $t_{min}$ & $t_{max}$ & $\sigma(M_X^2)$  & ${\mathcal L}/10^{38}$ &  HRS   & DVCS   & Time\\
(GeV$^2$)& (GeV)  &     & (GeV)   & (m)   & (deg)   & (deg)    & (GeV$^2$) &(GeV$^2$)  & (GeV$^2$)      & (cm$^{-2}$/s) &	(Hz)  & (Hz)   & (days)\\ \hline
3.0  & 6.6    &  0.36   &  4.35   & 1.5   & 11.7    & 7.1      & -0.16     & -0.42     & 0.23  &0.75   &   479    &   1.16&  3 \\
4.0  & 8.8    &  0.36   &  5.83   & 2.0   & 10.3    & 7.0      & -0.17     & -0.42     & 0.26  &1.3    &  842     &   1.74&  2 \\
4.55 &11.0    &  0.36   &  6.65   & 2.5   & 10.8    & 7.0      & -0.17     & -0.42     & 0.27  & 2     & 2460     &   4.63&  1 \\  \hline
3.1  & 6.6    &  0.5    &  3.11   & 1.5   & 18.5    & 11.0     & -0.37     & -0.64     & 0.17  &0.75   &  873     &   0.77&  5 \\
4.8  & 8.8    &  0.5    &  4.91   & 2.0   & 14.5    & 8.9      & -0.39     & -0.70     & 0.20  &1.3    &  716     &   0.82&  4 \\
6.3  &11.0    &  0.5    &  6.50   & 2.5   & 12.4    & 7.9      & -0.40     & -0.72     & 0.20  & 2.    &  778     &   0.99&  4 \\
7.2  &11.0    &  0.5    &  7.46   & 2.5   & 10.2    &  7.0     & -0.40     & -0.75     & 0.25  & 2.    &  331     &   0.53&  7 \\  \hline
5.1  & 8.8    &  0.6    &  4.18   & 1.5   & 17.8    & 10.4     & -0.65     & -1.06     & 0.16  &0.75   &  338     &   0.27& 13 \\
6.0  & 8.8    &  0.6    &  4.97   & 2.0   & 14.8    &  9.2     & -0.67     & -1.05     & 0.18  & 1.3   &  227     &   0.22& 16 \\
7.7  &11.0    &  0.6    &  6.47   & 2.5   & 13.1    &  8.6     & -0.69     & -1.10     & 0.20  & 2.    &  274     &   0.28& 13 \\
9.0  &11.0    &  0.6    &  7.62   & 3.0   & 10.2    &  7.3     & -0.71     & -1.14     & 0.22  & 3.    &  117     &   0.17& 20 
\end{tabular}
\end{ruledtabular}
\end{table}

\newcounter{stat}
\setcounter{stat}{250}
The detailed $(e,e')$ kinematics, calorimeter configuration, 
H$(e,e'\gamma)X$ missing mass resolution in $M_X^2$, count-rates, and beam 
time are summarized in Table~\ref{tab:Trues}. The beam time is chosen to 
give a total statistics of \thestat K events per $(\xBj,Q^2)$ setting.

\subsection{Systematic (Instrumental) Errors}
\newcommand{\previous}{E00-110}
Table \ref{tab:systematics} shows the main systematic errors on the 
cross-sections extraction during 
estimation of those errors for the proposed experiment. The main 
improvements of the errors are due to :
\begin{itemize}
\item[-] The beam polarization measurements: the upgrade of the Compton 
polarimeter along the Hall A beam line will reduce the relative precision 
$\Delta P/P$ from 2\% to 1\%. The average beam polarization for 
\previous{} was 75\% and is expected to be 85\% for the proposed 
experiment.
\item[-] Drift chamber multi-tracks : the number of events for which 
multiple tracks are reconstructed in the focal plane of the HRS is 
proportional to the single rates. Multiple tracks events are discarded 
during the analysis and the live time of the experiment corrected 
accordingly. Because the proposed experiment is measured deeper in the DIS 
region, we expect that the correction will be smaller as well as its 
uncertainty.
\item[-] $\pi^0$ subtraction : the limitation on the precision of the 
correction for \previous{} was both statistical and limited by the 
kinematical range of the recorded $\pi^0$ events. For the proposed 
experiment, we plan to improve the trigger such that more $\pi^0$ events 
are recorded. A special trigger with lower threshold but the requirement 
of two clusters being above it, will be implemented.
\item[-] e(p,e'$\gamma$)$\pi$N contamination : the missing mass squared 
resolution is the key ingredient for this contamination. The proposed 
estimation takes into account the variation of this resolution for the new 
kinematics.
\end{itemize}
The systematic errors in Table~\ref{tab:systematics} are summarized 
separately for the cross section sum $d\overrightarrow{\sigma} + 
d\overleftarrow{\sigma}$, and for the cross section difference 
$d\overrightarrow{\sigma} - d\overleftarrow{\sigma}$.

\begin{table}
\begin{center}
\begin{tabular}{|l| l|c|c|}
\hline
\multicolumn{2}{|c}{Type }          & \multicolumn{2}{|c|}{ Relative errors  (\%)} \\
\multicolumn{2}{|c|}{}          & \previous & proposed \\ \hline \hline
Luminosity & target length and beam charge           & 1   & 1      \\ \hline
HRS-Calorimeter  & Drift chamber multi-tracks        & 1.5 & 1      \\
     & Acceptance                                    & 2   & 2      \\                   
     & Trigger dead-time                             & 0.1 & 0.1    \\  \hline
DVCS selection & $\pi^0$ subtraction                 & 3   & 1      \\ 
               & e(p,e'$\gamma$)$\pi$N contamination & 2   & 3      \\
               & radiative corrections               & 2   & 1      \\ \hline 
\multicolumn{2}{|c|}{Total cross section sum}        & 4.9 & 4.1    \\ \hline \hline
 Beam    & Polarization $\Delta P/P$                 & 2   & 1      \\ \hline
\multicolumn{2}{|c|}{Total cross section difference} & 5.3 & 4.2    \\ \hline
\end{tabular}
\caption{\label{tab:systematics} Relative systematic error budget for \previous{}
 and for the proposed experiment. 
}
\end{center}
\end{table}

\subsection{Statistics}

\begin{figure}[ht]
\includegraphics[width=0.49\linewidth]{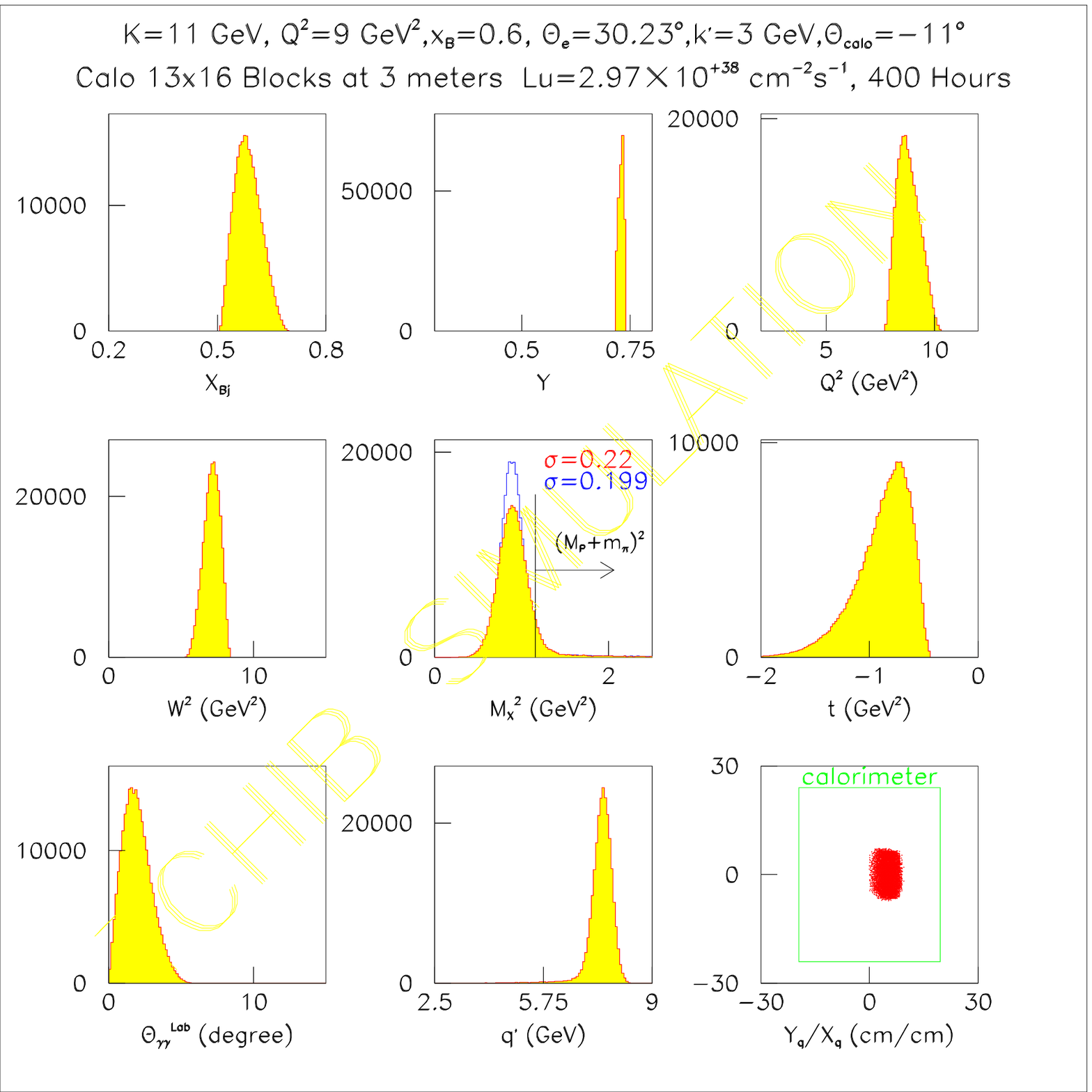}\hfill
\includegraphics[width=0.49\linewidth]{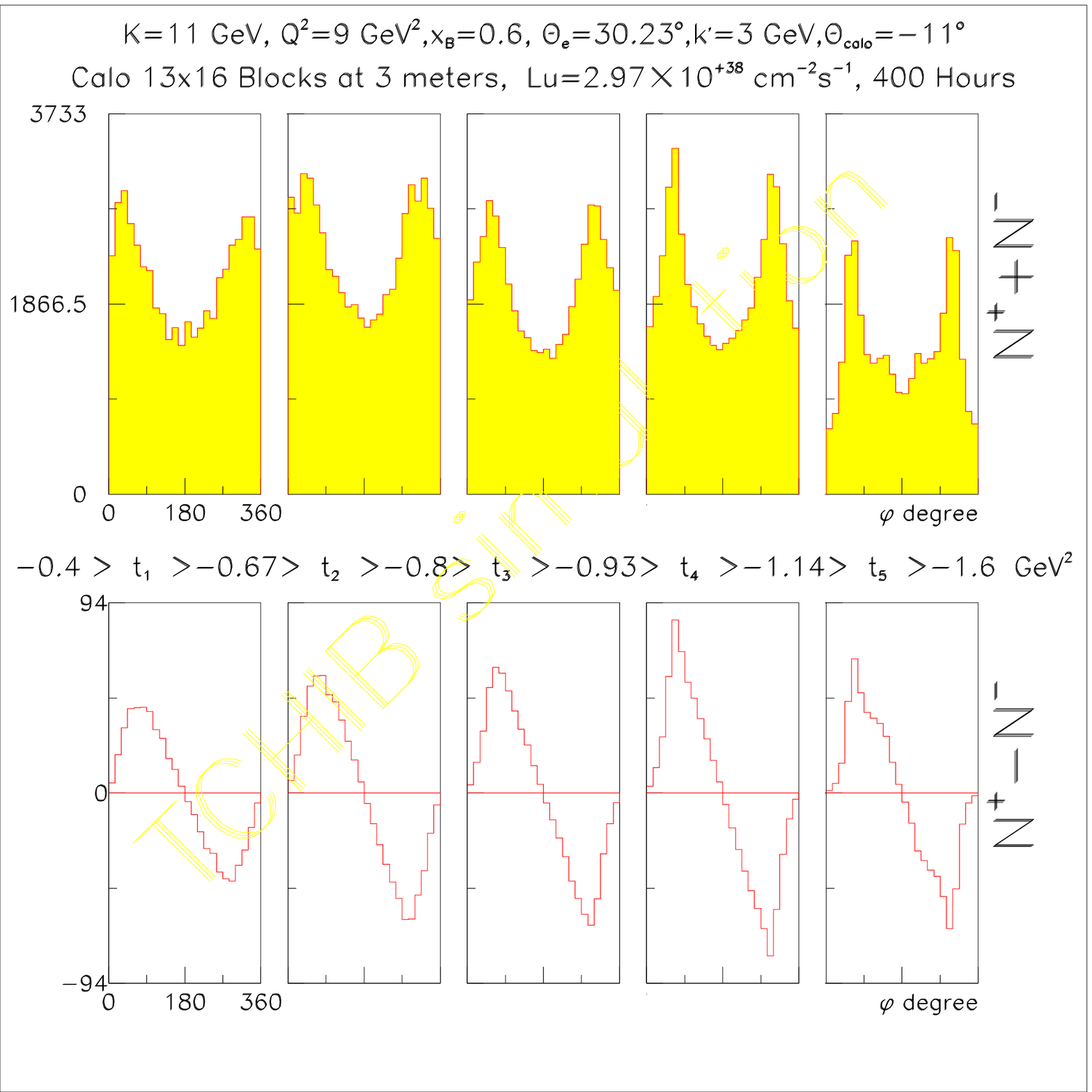}
\begin{minipage}{0.45\linewidth}
\caption[DVCS Kinematics]{\label{fig:kin69} DVCS Distributions for setting 
$k=11$ GeV, $Q^2 = 9$ \GeVsq, $\xBj=0.6$.\\ Top Left:  Cross section 
weighted acceptance distributions.\\ Top Right:  Helicity sum and helicity 
difference projected counts as a function of $\phi_{\gamma\gamma}$ in five 
bins in $t$. \\ Bottom Right:  Helicity sum and helicity difference 
projected cross sections, with statistical uncertainties, as functions of 
$\phigg$ in the same bins in $t$.}
\end{minipage} \hfill
\begin{minipage}{0.49\linewidth}
\includegraphics[width=\linewidth]{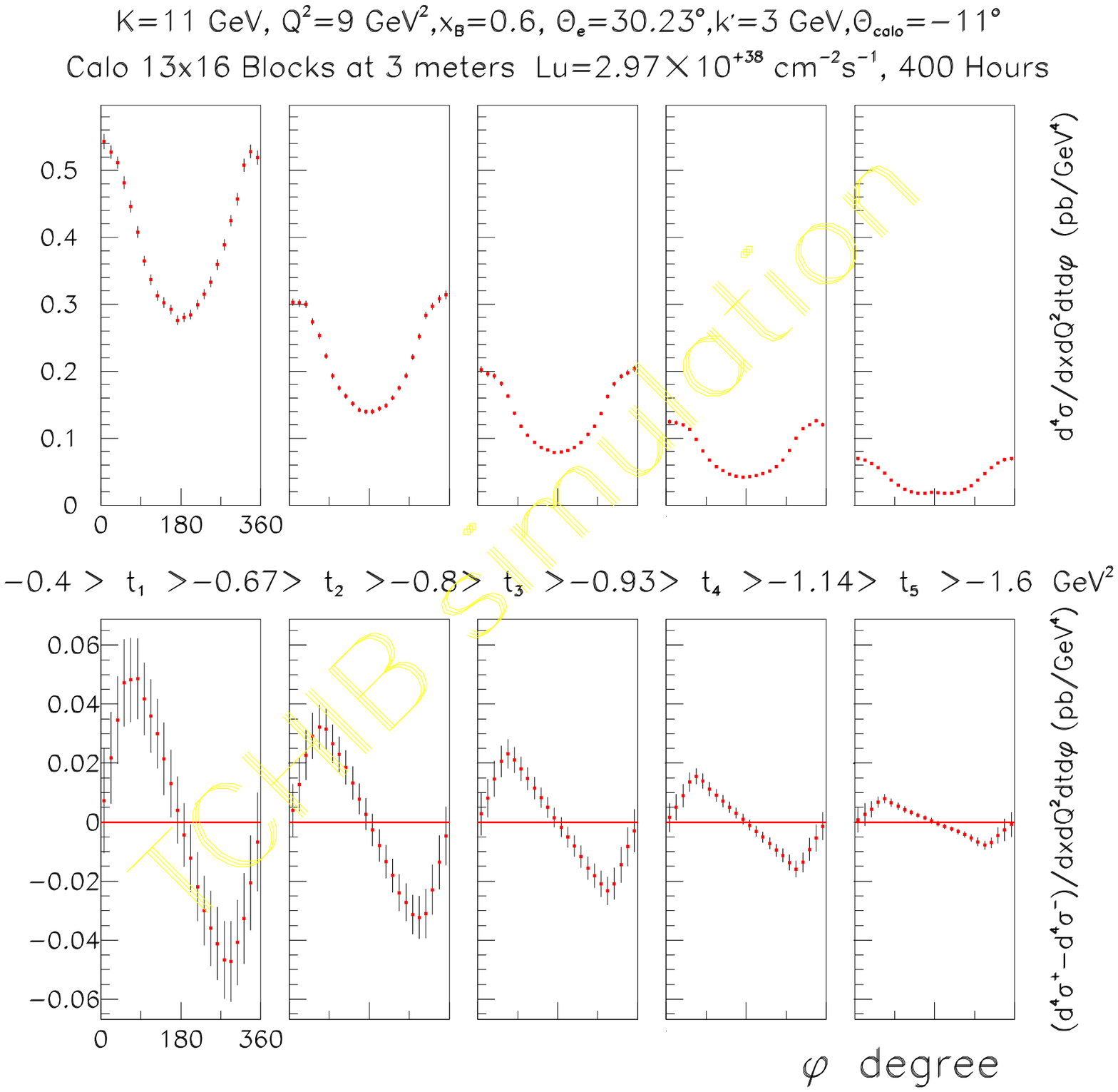}
\end{minipage}\\[1em]
\hrulefill
\end{figure}

\begin{figure}[ht]
\includegraphics[width=0.49\linewidth]{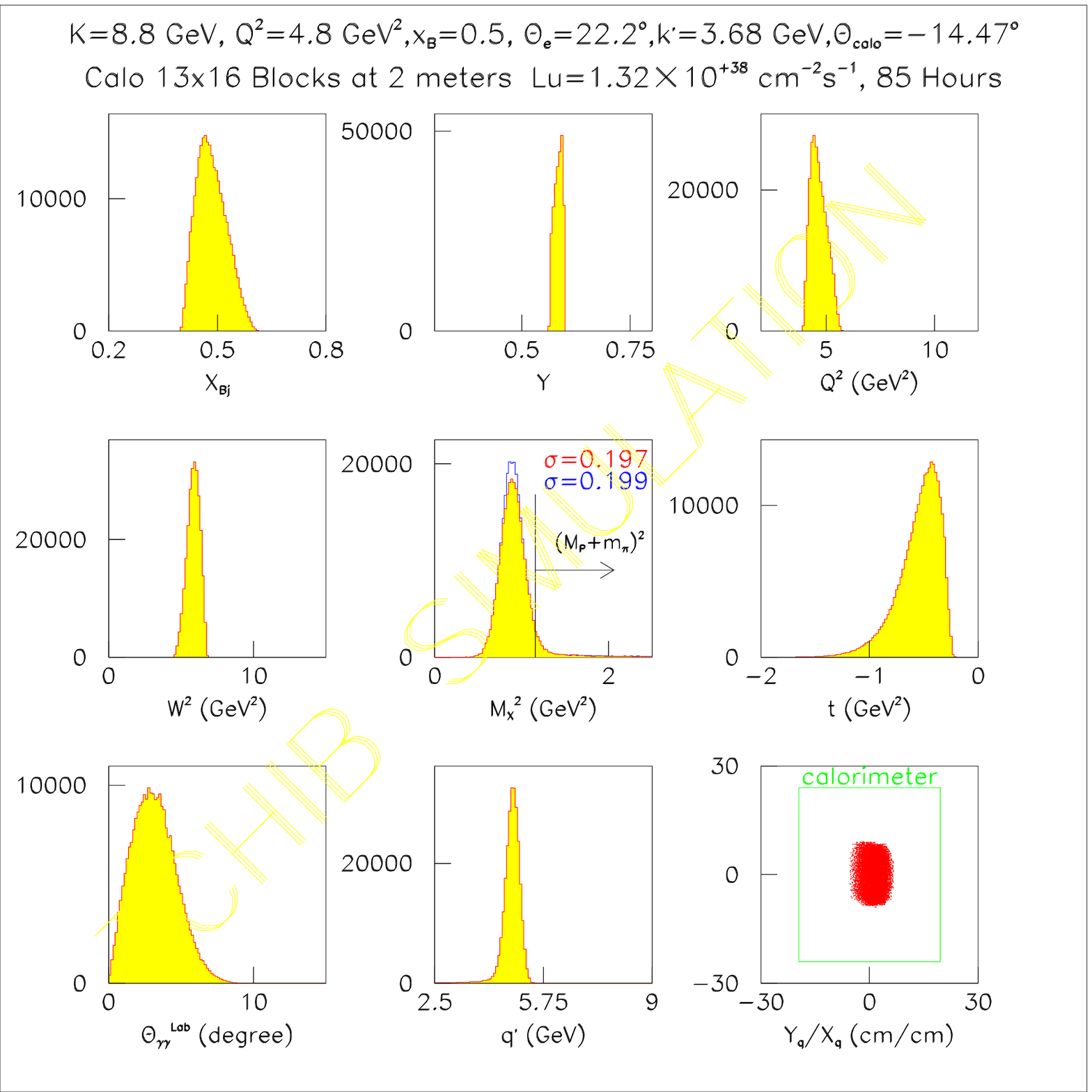}\hfill
\includegraphics[width=0.49\linewidth]{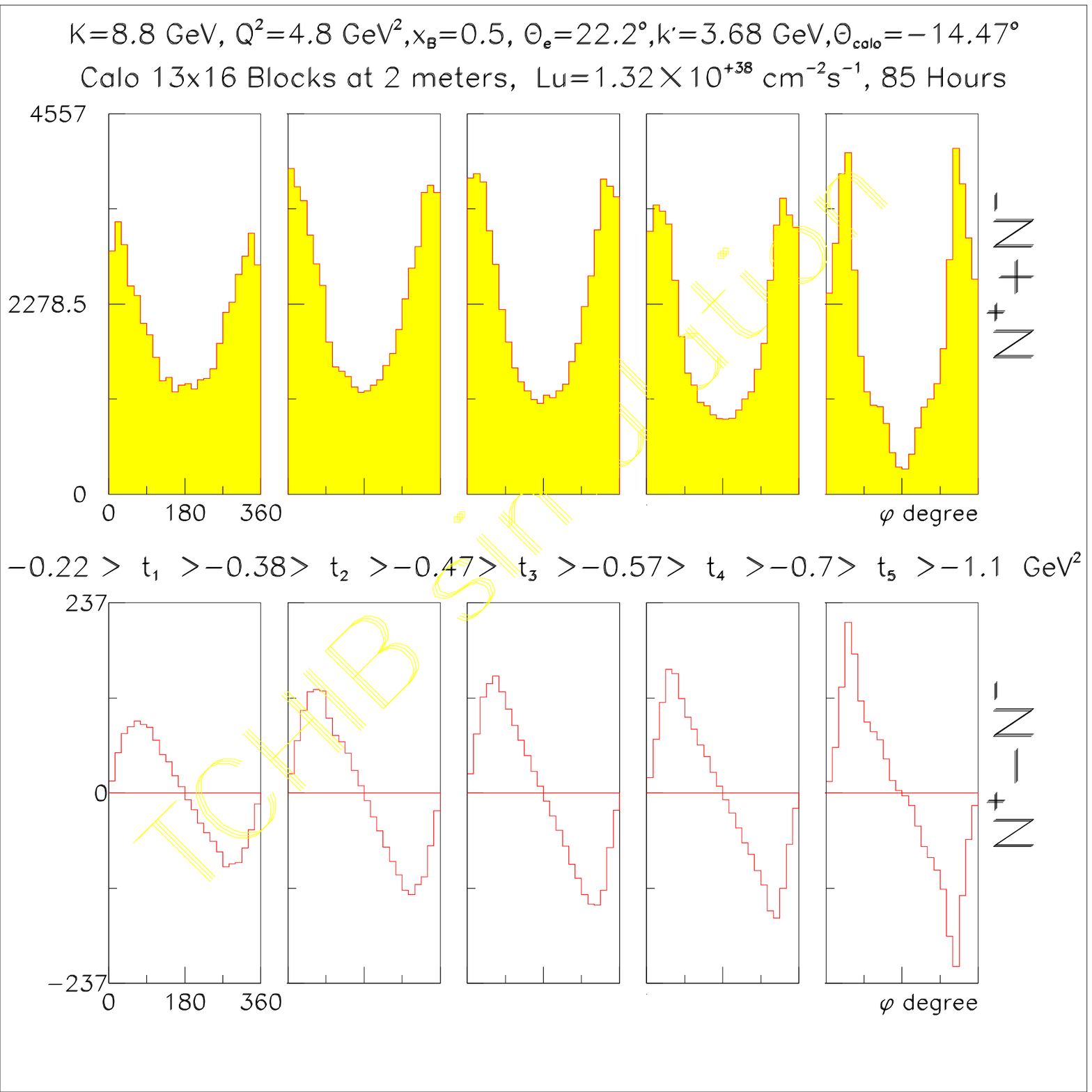}
\begin{minipage}{0.45\linewidth}
\caption[DVCS Kinematics]{\label{fig:kin88} DVCS Distributions for setting 
$k=8.8$ GeV, $Q^2 = 4.8$ \GeVsq, $\xBj=0.5$.\\ Top Left:  Cross section 
weighted acceptance distributions.\\ Top Right:  Helicity sum and helicity 
difference projected counts as a function of $\phi_{\gamma\gamma}$ in five 
bins in $t$. \\ Bottom Right:  Helicity sum and helicity difference 
projected cross sections, with statistical uncertainties, as functions of 
$\phigg$ in the same bins in $t$.}
\end{minipage} \hfill
\begin{minipage}{0.49\linewidth}
\includegraphics[width=\linewidth]{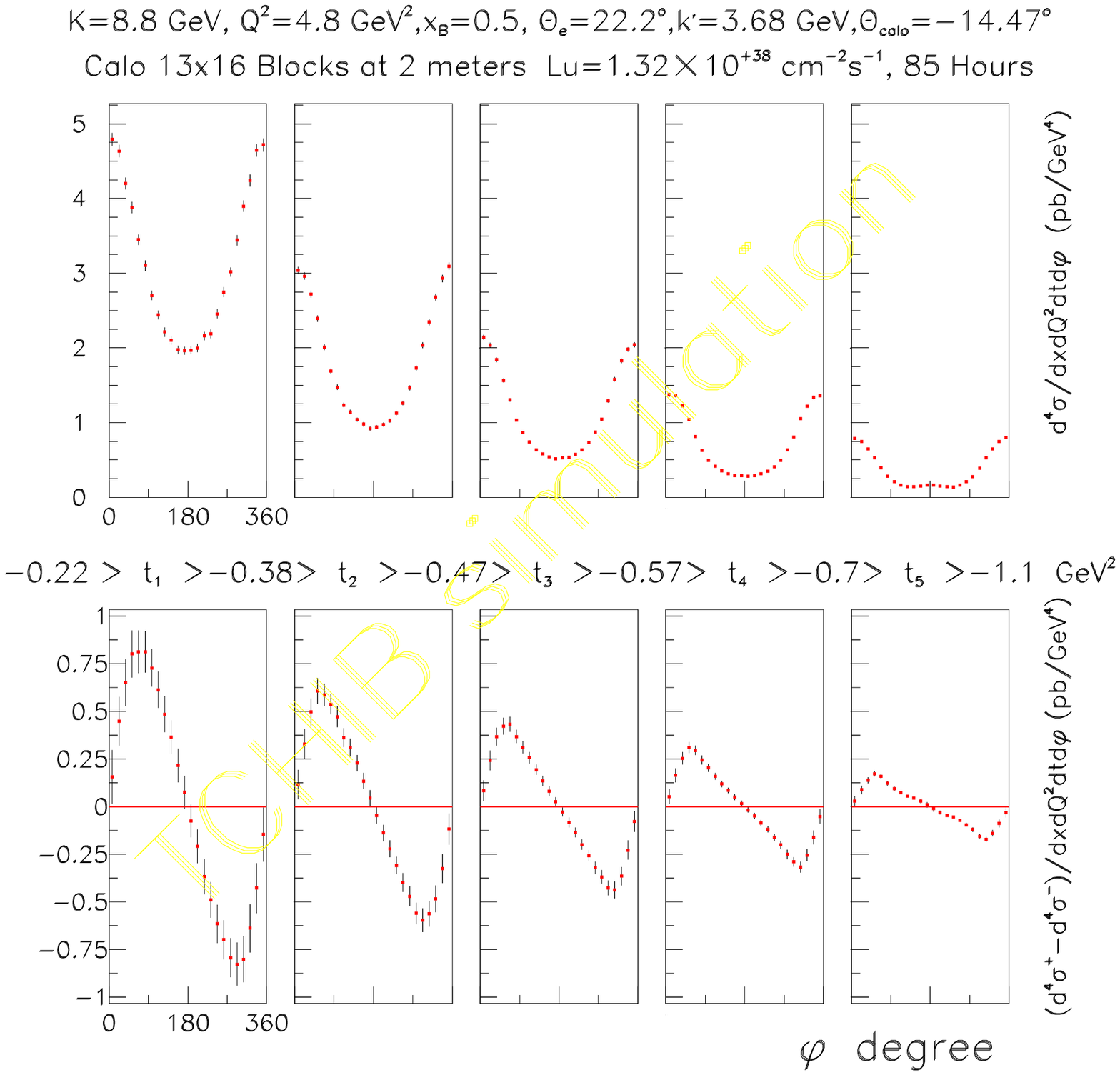}
\end{minipage}\\[1em]
\hrulefill
\end{figure}

\begin{figure}[ht]
\includegraphics[width=0.49\linewidth]{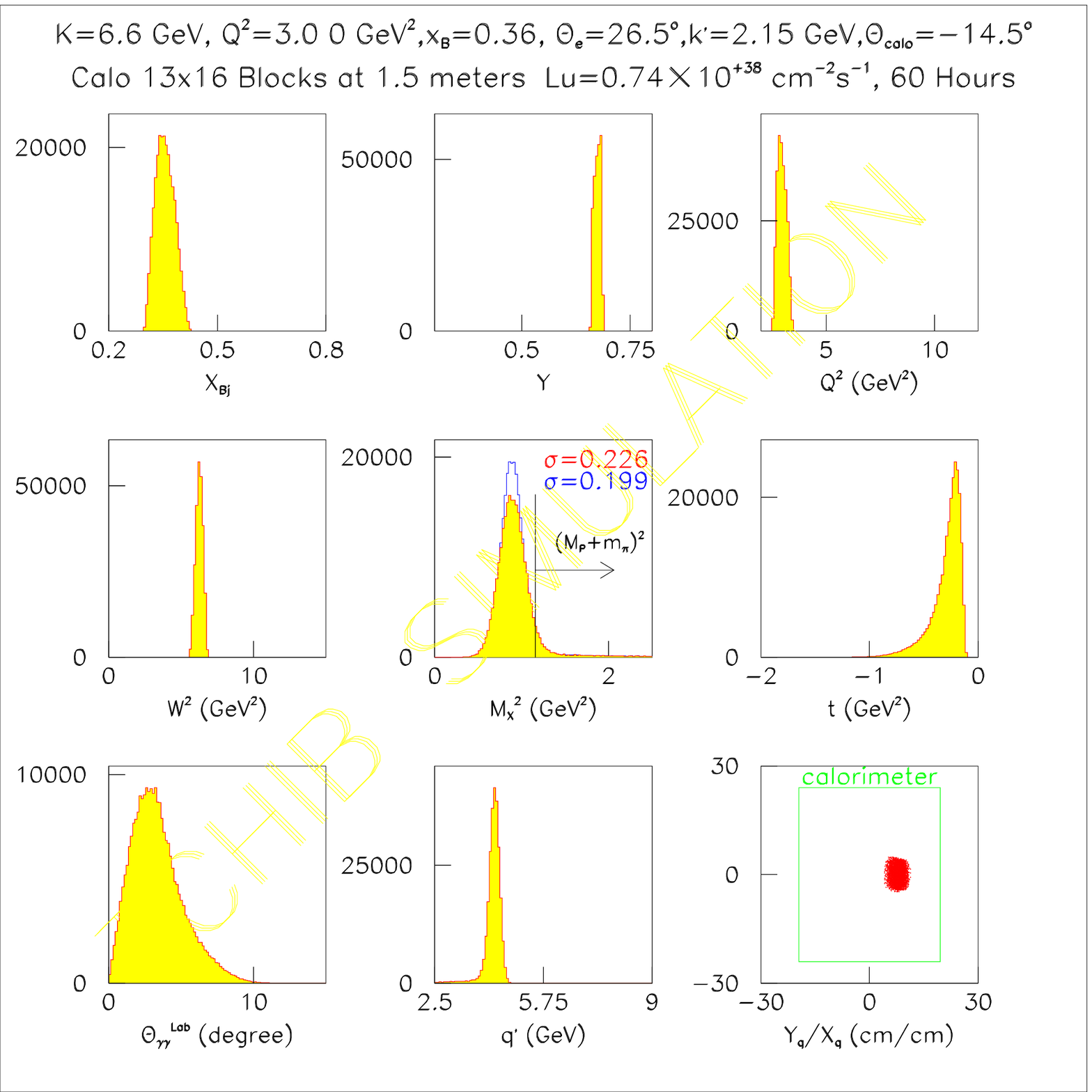} \hfill
\includegraphics[width=0.49\linewidth]{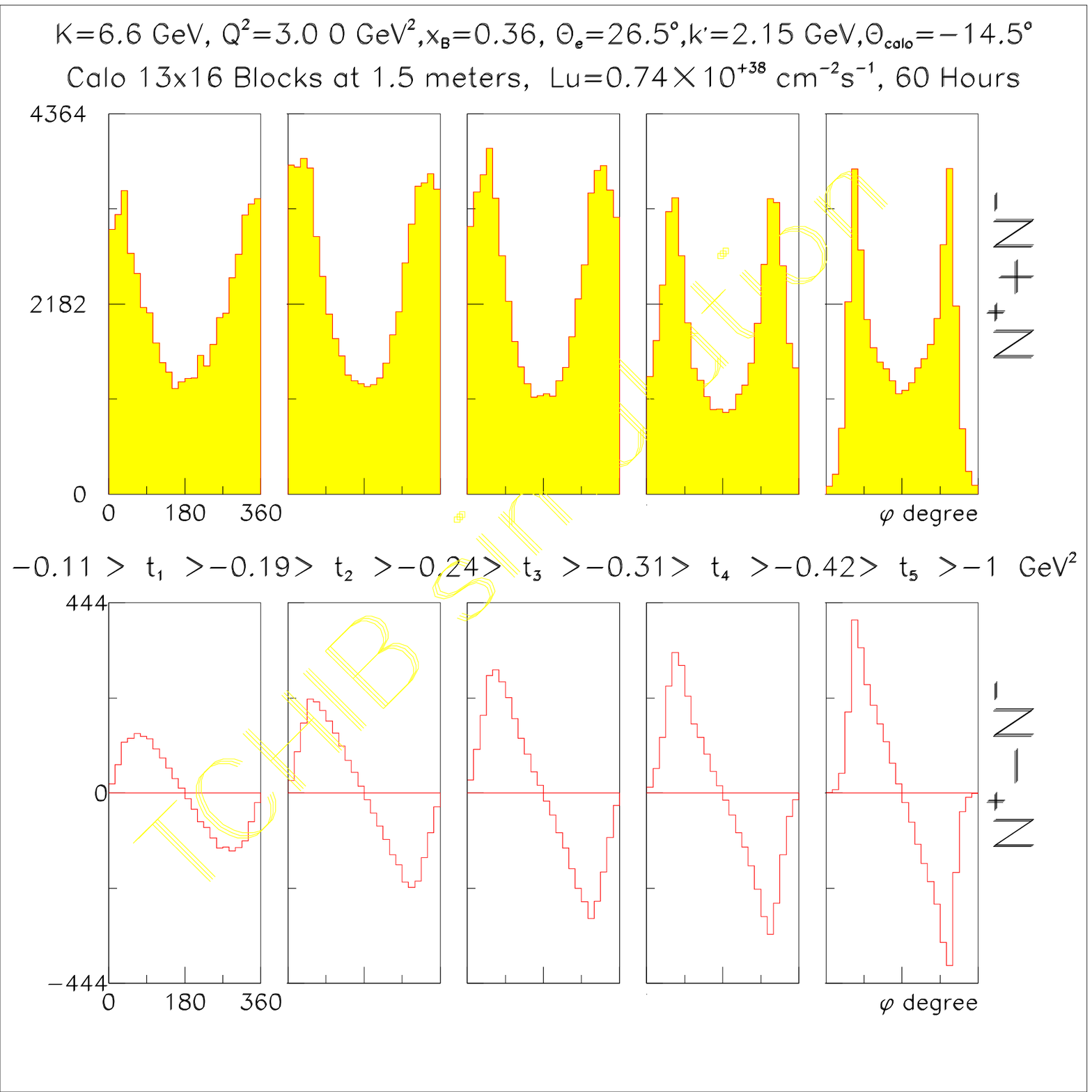}
\begin{minipage}{0.45\linewidth}
\caption[DVCS Kinematics]{\label{fig:kin66} DVCS Distributions for setting 
$k=6.6$ GeV, $Q^2 = 3.0$ \GeVsq, $\xBj=0.36$.\\ Top Left:  Cross section 
weighted acceptance distributions.\\ Top Right:  Helicity sum and helicity 
difference projected counts as a function of $\phi_{\gamma\gamma}$ in five 
bins in $t$. \\ Bottom Right:  Helicity sum and helicity difference 
projected cross sections, with statistical uncertainties, as functions of 
$\phigg$ in the same bins in $t$.}
\end{minipage} \hfill
\begin{minipage}{0.49\linewidth}
\includegraphics[width=\linewidth]{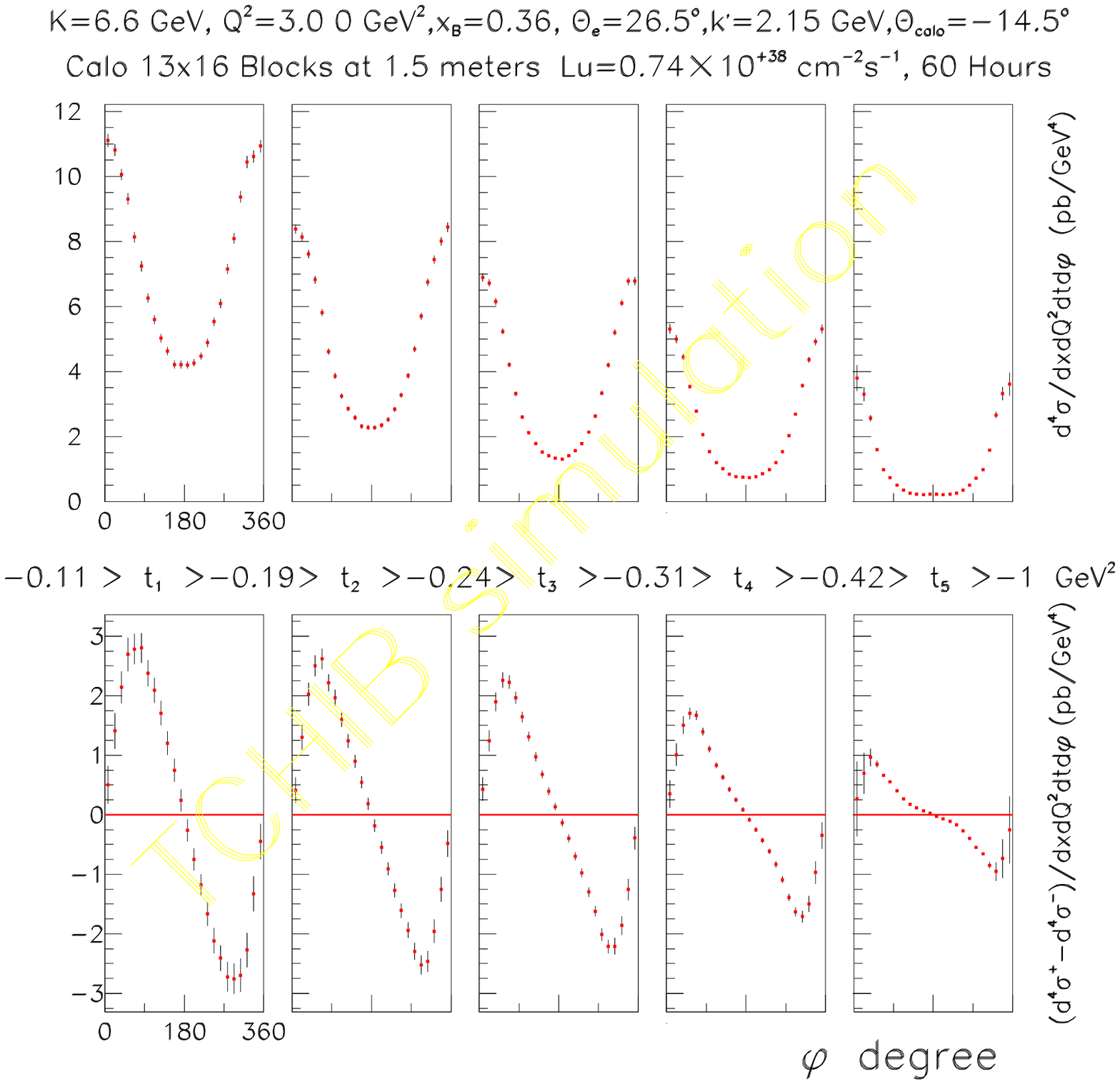}
\end{minipage}\\[1em]
\hrulefill
\end{figure}

Fig.~\ref{fig:kin69}--\ref{fig:kin66} show acceptance, counting 
statistics, and cross section distributions for specific $(\xBj,Q^2)$ 
kinematics. In each figure, we present the cross section weighted 
acceptance distribution of a basic set of kinematic variables in the upper 
left panel. The central plot of this panel is the H$(e,e'\gamma)X$ missing 
mass squared $M_X^2$ distribution for simulated exclusive events.  The 
blue curve is a simulation for E00-110.  The missing mass resolution 
$\sigma(M_X^2)$ for each setting is tabulated in Table~\ref{tab:Trues}. 
For E00-110, the resolution is $\sigma(M_X^2)= 0.20$ GeV$^2$. In the lower 
right kinematic plot, The green rectangle is the fiducial surface of 
$11\times 14$ blocks of the calorimeter (we require the centroid of EM 
shower to be a a minimum of of one block from the calorimeter edge). The 
points in red on this plot show the distribution of directions of the 
${\bf q}$-vector. The upper right panel of the figures shows the counting 
statistics (helicity sum and helicity difference) as functions of $\phigg$ 
for five bins in $t$.  As $-t$ increases from left to right, the 
simulation shows the gradual loss of acceptance for $\phigg$ near 
$0^\circ$, due to the off-centered calorimeter. The bottom right panel 
shows the projected helicity sum and helicity difference cross sections, 
with statistical errors, in the same bins in $t$ and $\phigg$.  The 
statistics are for the projected beam time of this proposal.  These 
estimates were made with our simulation code TCHIB.  The cross section 
model for these estimates is described in \S \ref{sec:VGG}.

With \thestat K events in each $(\xBj,Q^2)$ setting, we obtain a high 
precision determination of all the observables detailed in the Section 
\ref{sec:BHDVCS}.

\clearpage

\subsection{\label{sec:VGG}Illustration of Prospective Physics}

In each setting $(\xBj,Q^2)$ setting, we will extract in the $t$-bins 
defined in Fig.~\ref{fig:kin69}--\ref{fig:kin66}, the experimental Twist-2 
and Twist-3 BH$\cdot$DVCS interference observables defined in 
Section.~\ref{sec:BHDVCS}. In addition, we will obtain upper bounds on the 
gluon transversity terms $\cos(3\phigg)$ and $\sin(3\phigg)$.

\begin{figure}[ht]
\includegraphics[width=0.9\linewidth]{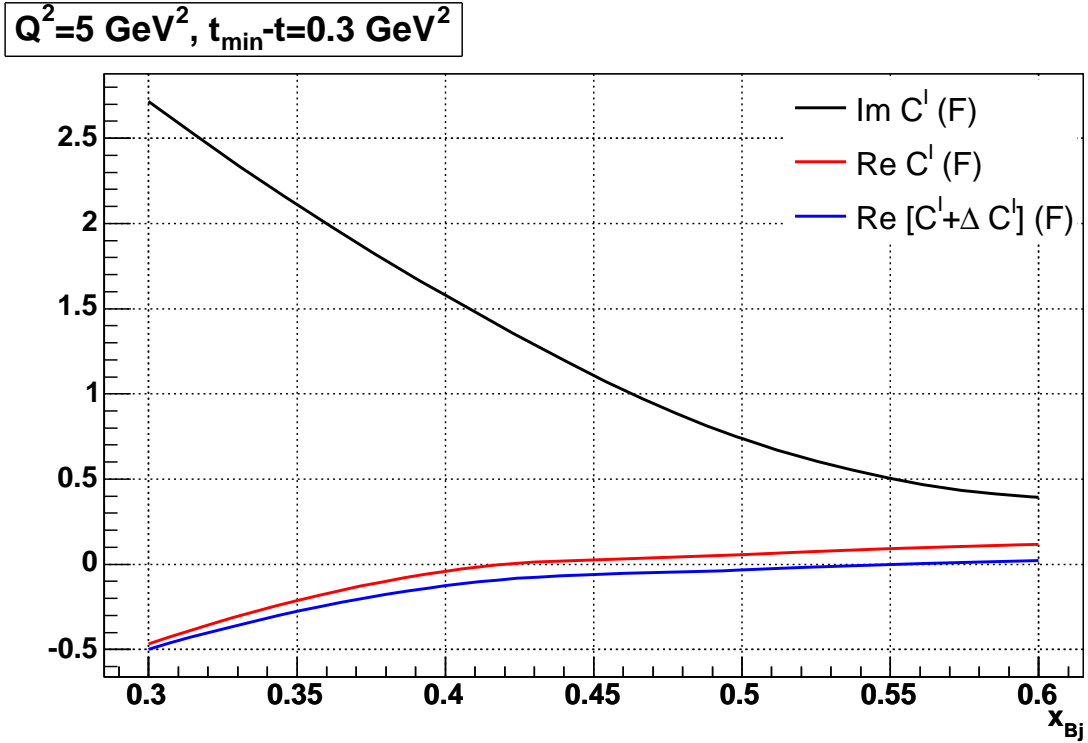}
\caption{\label{fig:vsx} The three Twist-2 observables, as functions of 
$\xBj$ at $t_{\text min}-t=0.3$ GeV$^2$.  The model is from VGG 
\cite{Vanderhaeghen:1999xj,Goeke:2001tz,Guidal:2004nd}.
}
\end{figure}

\begin{figure}[ht]
\includegraphics[width=0.8\textwidth]{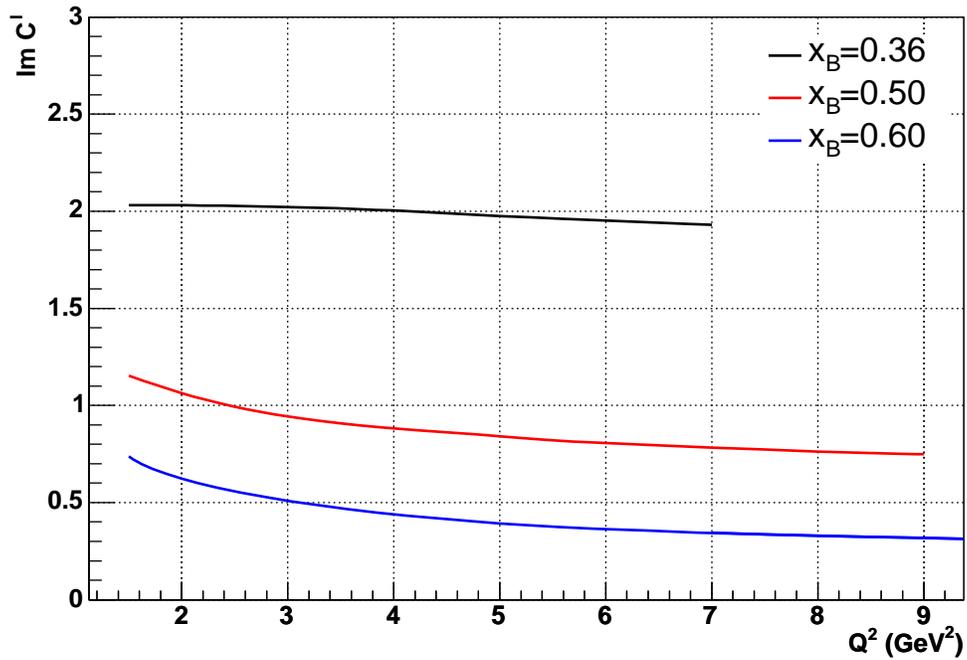}
\caption{\label{fig:c1QCDevolution} QCD evolution of $\mathcal C^{\mathcal 
I}$, as obtained from DGLAP evolution of $H(x,\xi , t ;Q^2)$ in the VGG 
model \cite{Vanderhaeghen:1999xj,Goeke:2001tz,Guidal:2004nd}.}
\end{figure}

We illustrate the possible $\xBj$-dependence of the Twist-2 coefficients 
in Fig.~\ref{fig:vsx}.  The principal-value integrals in the $\Real 
\mathcal C^{\mathcal I}$ and $\Real\Delta\mathcal C^{\mathcal I}$ must 
change sign somewhere in the range $0\le\xi<1$. In 
Fig.~\ref{fig:c1QCDevolution}, we present the impact on $\Imag[\mathcal 
C^{\mathcal I}]$ of the DGLAP evolution of $H(x,\xi,t;Q^2)$ in the VGG 
model \cite{Vanderhaeghen:1999xj,Goeke:2001tz,Guidal:2004nd}. The VGG 
model uses the double distributions for $H$:
\begin{eqnarray}
H_f(x,\xi,t) &=& \int d\beta d\alpha \delta(x-\beta-\alpha\xi) F_f(\beta,\alpha,t) \label{eq:DD}\\
F_f(\beta,\alpha,t) &=& \left. h(\beta,\alpha)q_f(\beta)
   \right/ \beta^{\alpha' t} \label{eq:Regge}\\
h(\beta,\alpha) &=&
C(b) \left[(1-|\beta|)^2-\alpha^2\right]^b\left/\left(1-|\beta|\right)^{2b+1}\right. \label{eq:Profile}
\end{eqnarray}
The limits of integration in Eq.~\ref{eq:DD}, and the normalization 
constant $C(b)$ are defined in \cite{Goeke:2001tz}.  In these illustrative 
plots, we have taken the profile parameter $b=1$ and the Regge parameter 
$\alpha' = 0.8$ GeV$^{-2}$.  A similar form is used for $\tilde H$, with 
$q_f(\beta)$ replaced with $\Delta q_f(\beta)$.  We also set $E=0$ in 
these figures. For our count rate estimates in the previous section, we 
have used the factorized ansatz which is obtained in the $b\rightarrow 
\infty$ limit. In Fig.~\ref{fig:c1QCDevolution}, the DGLAP evolution is 
applied only to the parton distributions $q_f(\beta)$. We note that the 
experimental contribution of the Twist-3 term $\langle \eta_1\rangle 
\Imag[\mathcal C_1^{\rm DVCS}]$ (Eq.~\ref{eq:ImC1exp}) will decrease in 
proportion to $1/\sqrt{Q^2}$. At the same time, higher twist contributions 
in $\Imag[\mathcal C^{\mathcal I}]$ will decrease in proportion to $1/Q^2$ 
(or higher powers).  Thus it should be possible to extract the Twist-2 and 
Twist-3 contributions from the $Q^2$-dependence, with only a small model 
dependence from the distinct QCD evolution of the three contributing GPDs.

\begin{figure}[hb]
\includegraphics[width=0.8\textwidth]{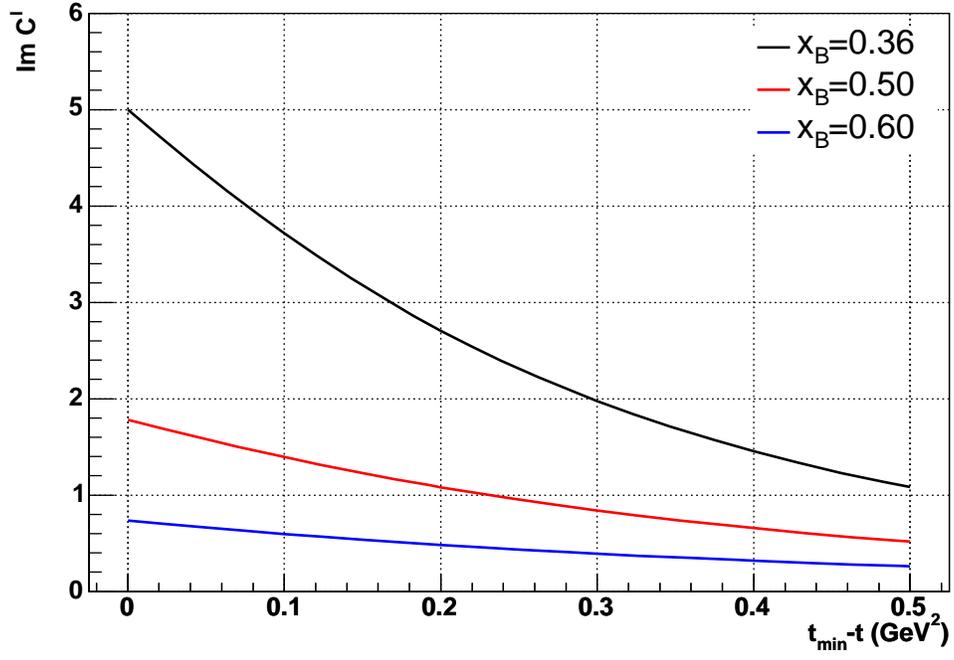}
\caption{\label{fig:t-dependence} Dependence of $\Imag\mathcal C^{\mathcal 
I}(\mathcal F)$ as a function of $t_{\text min}-t$ for the three values of 
$\xBj$ of this proposal at constant $Q^2$. The curves are obtained from 
the contributions of $H(x,\xi , t )$ and $\tilde{H}(x,\xi,t)$ in the VGG 
model \cite{Vanderhaeghen:1999xj,Goeke:2001tz,Guidal:2004nd}.}
\end{figure}

Fig.~\ref{fig:t-dependence} illustrates the $t$-dependence of 
$\Imag\mathcal C^{\mathcal I}(\mathcal F)$, plotted here as a function of 
$t_{\text min}-t$ for our three values of $\xBj$.  The $t$-dependence of 
this observable is the bilinear combination of the Compton form factors 
with the proton form factors in the BH amplitude.

\clearpage
\section{\label{sec:Summary}Summary}

With a high precision spectrometer, and a compact, hermetic 
high-performance calorimeter in Hall A, we can obtain high statistics 
absolute measurements of $\vec{e}p\rightarrow e p \gamma$ cross sections 
in a wide DVCS kinematic domain with CEBAF at 12 GeV.  We expect to obtain 
precision measurements of three Twist-2 DVCS observables and two Twist-3 
DVCS observables. The $Q^2$ dependence will provide stringent tests of the 
factorization theorem, and quantify the contribution of higher twist terms 
(which can be modelled as the hadronic content of the photon). The 
$t$-dependence as a function of $\xBj$ (or $\xi$) of the observables will 
provide our first study of transverse profile of the proton as a function 
of quark light-cone momentum fraction.  We note that in this kinematic 
region, models such as the VGG model presented in \S \ref{sec:VGG} must be 
considered uncertain by at least a factor of two. Our separations of the 
$\Real$ and $\Imag$ parts of the DVCS observables will provide a 
calibration for present and future measurements of relative asymmetries.  
Through the principle value integrals, the $\Real$ part provides access to 
regions of $(x,\xi)$-space that are not directly accessible via the 
$\Imag$ part of the GPDs.  The $\Imag$ part observables restrict the GPDs 
to the sum of the points $x=\pm\xi$.

We will obtain precision measurements of the $\sigma_L + 
\sigma_T/\epsilon_L$, $\sigma_{LT}$, $\sigma_{TT}$ and $\sigma_{LT'}$ 
cross sections of neutral pion electro-production H$(e,e'\pi^0)p$.  The 
factor $\epsilon_L$ is the degree of longitudinal polarization of the 
virtual photon..  The transverse cross section $\sigma_T$ is expected to 
fall faster with $Q^2$ (by one power of $1/Q^2$) than the leading twist 
term in the longitudinal cross section $d\sigma_L$.  Therefore, from the 
$Q^2$ dependence of the H$(e,e'\pi^0)p$ cross section, we can extract the 
leading twist (handbag) contribution to this process.

We require 88 days of production running, with an anticipated additional 
12 days interlaced for optical curing of the calorimeter.  This will 
provide a major survey of proton DVCS in almost the entire kinematic range 
accessible with CEBAF at 12 GeV.
\clearpage

\centerline{\large \bf REFERENCES}
\bibliography{Compton}
\clearpage

\appendix
\section{Contributions to Hall A Equipment for 11 GeV}

Our committments to the Hall A Base Equipment are detailed in 
Table~\ref{tab:equip}. The collaboration will also seek additional funds 
for the upgrade to the calorimeter and beam line.

\begin{table}[h]
\caption{\label{tab:equip} Contributions of DVCS Collaboration to 
Hall A  Base Equipment}
\begin{ruledtabular}\begin{tabular}{lclr}
Institution & Funding Source & \multicolumn{1}{c}{Project Description} & Equipment \\ \hline
LPC-Clermont-Ferrand & IN2P3-CNRS & Compton Polarimeter: & \$125K\\ 
                     &            & HRS Trigger Electronics & \$39K \\ \hline
\multicolumn{3}{l}{Technical Manpower for Base Equipment} & (months) \\ \hline
LPC-Clermont-Ferrand & IN2P3-CNRS & Compton Polarimeter: & \\
                     &            &  \multicolumn{1}{r} Engineer\ . & 24\\
                     &            &  \multicolumn{1}{r} Technician & 24\\
                     &            &  \multicolumn{1}{r} Physicist\ . & 12\\
                     &            & HRS Trigger Electronics & \\
                     &            &  \multicolumn{1}{r} Engineer\ . & 6\\
                     &            &  \multicolumn{1}{r} Technician & 6\\
Old Dominion U. & DOE  & HRS DAQ \hfill Faculty & 4 \\
                     & (ongoing grant) & PostDoc & 12 \\
\end{tabular}
\end{ruledtabular}
\end{table}

\section{Preparation of Extensions}

We are actively preparing a number of extensions of this proposal, which 
will lead to a comprehensive program of measurements exploiting the unique 
capabilities of Hall A.

\subsection{Deuterium}

\begin{figure}
\includegraphics[width=\linewidth]{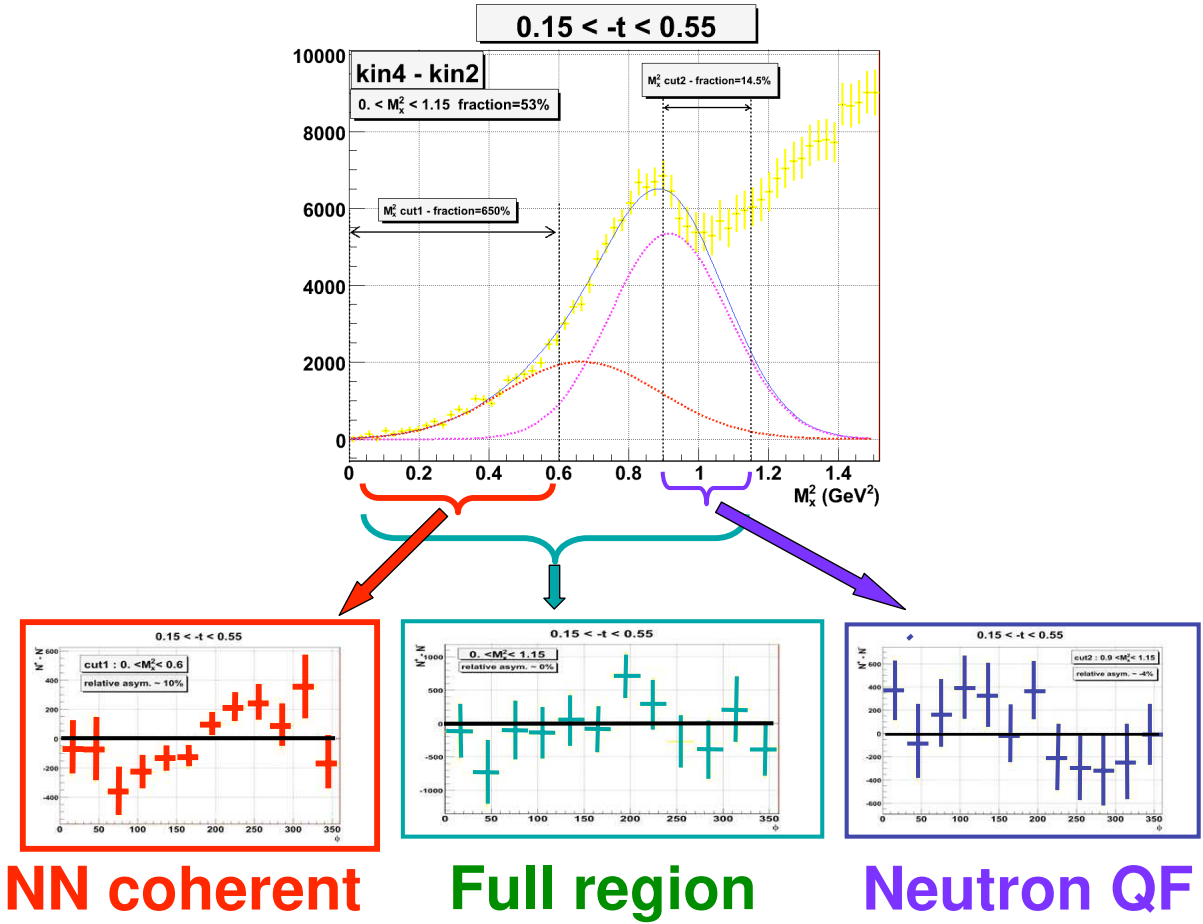}
\caption{\label{fig:Neutron} Top: Missing mass distribution of E03-106 
D$(e,e'\gamma)X$ events, after subtracting a normalized ensemble of 
H$(e,e'\gamma)X$ events. The kinematics are $Q^2=1.9$ GeV$^2$, 
$\xBj=0.36$, and $\langle t\rangle=-0.3$ GeV$^2$. The missing mass is 
calculated for $N(e,e'\gamma)N$ kinematics. Thus the quasi-free 
$n(e,e'\gamma)n$ peak appears at $M_X^2 = M^2$ and and the coherent 
D$(e,e'\gamma)$D peak is broadened by the $t$-acceptance of the events, 
and appears at $M_X^2 = M^2+t/2$.  In the region $M_X^2>M^2+t/2$, there 
can also be a contribution from D$(e,e'\gamma)np$ on high momentum 
correlations in the initial state of the Deuteron ('$NN$-Coherent').  The 
curves are a two-gaussian fit, with the relative positions constrained to 
$t/2$. The bottom three plots show the helicity weighted statistics 
$N(+)-N(-)$ in the three missing mass regions: $0.0<M_X^2<0.6$ GeV$^2$ 
(left); $0.0<M_X^2<1.15$ GeV$^2$ (center); and $0.9<M_X^2<1.15$ GeV$^2$ 
(right).  The '$NN$-coherent' and 'Quasi-Free Neutron' regions have 
opposite asymmetry.}
\end{figure}

We currently have preliminary quasi-free and coherent DVCS data from Hall 
A E03-106 in the channels D$(e,e'\gamma)pn\oplus{\rm D}(e,e'\gamma)D$. 
These two channels are (partially) separated experimentally by the 
differential recoil of the coherent deuteron.  In a plot of $M_X^2$, 
calculated for H$(e,e'\gamma)p$ kinematics, the coherent Deuteron peak 
appears at $M_X^2=M_p^2+t/2$ (Fig.~\ref{fig:Neutron}).  The smearing of 
the quasi-elastic D$(e,e'\gamma)pn$ events by the internal momentum 
distribution of the $np$ wavefunction of the deuteron is less than our 
experimental resolution.  Thus our isolation of the D$(e,e'\gamma)pn$ 
channel from the inelastic D$(e,e'\gamma)NN\pi$ channel is only slightly 
degraded compared to the identification of the H$(e,e'\gamma)p$ exclusive 
channel.

In a Quasi-Free (QF) model, the neutron DVCS cross section can be obtained 
from the following subtraction:
\begin{alignat}{1}
d^5\sigma(\vec{e}n\rightarrow e'\gamma n) = & 
d^5\sigma(\vec{e}D\rightarrow e'\gamma np) - d^5\sigma(\vec{e}p\rightarrow e'\gamma p) 
\label{eq:Neutron}
\end{alignat}
All observables on the neutron have the same form as for the proton.  
Using isospin symmetry, the neutron observables interchange the role of up 
and down quarks. Whereas the proton cross sections are four times more 
sensitive to up than down quarks, the reverse is true for the neutron.  
As a consequence, based on the forward parton limits of the GPDs and the 
Form Factor constraints of the first moments of the GPDs, we expect the 
neutron observable $\mathcal C^{\mathcal I}$ to be dominated by the GPD 
$E$.

In Fig.~\ref{fig:Neutron}, we show the helicity correlated statistics in 
three zones in missing mass of the 'neutron' spectrum of 
Eq.~\ref{eq:Neutron}. We note that the signal obtained in the 
'$NN$-coherent' region $M_X^2<0.6$ GeV$^2$ is of opposite sign from the 
signal in the Quasi-Free region near $M_X^2=M^2$.

We expect to prepare a separate proposal for $D(e,e'\gamma)$ measurements 
for the next 12 GeV PAC. We expect the Deuterium proposal to include a 
subset of the present kinematics. We would propose to run the two 
experiments concurrently, with alternating sequences of Hydrogen and 
Deuterium data taking, to minimize systematic errors.

\subsection{Recoil Polarimetry}

A full DVCS program requires proton polarization measurements. and 
detection of nuclear recoils in coherent $A(e,e\gamma A)$ DVCS reactions. 
The single- and double-spin observables of proton recoil polarization in 
$\vec{e} p \rightarrow e' \vec{p} \gamma $ and functionally equivalent to 
the observables for polarized targets: $\vec{e} \vec{p} \rightarrow e' p 
\gamma$ (D. M\"uller, private communication 2006). We are working on a 
conceptual design for a large acceptance recoil polarimeter.  With an 
analyzer of 7 cm C, we can achieve a Figure of Merit $\ge 0.005$ 
(scattering probability times analyzing power squared) for $p_p>550$ 
MeV/c. This could measure both longitudinal and transverse recoil proton 
polarization in DVCS. This would operate in an axial magnetic field at 
high luminosity in Hall A, in conjunction with the equipment of this 
proposal. We intend to have a proposal ready for PAC 32. (Summer 2007).

\end{document}